\documentclass[acmsmall]{acmart}
\usepackage{algorithmic}
\usepackage[ruled,linesnumbered]{algorithm2e}
\usepackage{graphicx}
\usepackage{subfigure}

\usepackage{color}
\usepackage{xspace}
\newcommand{\mypara}[1]{{\noindent\textbf{#1}}}

\AtBeginDocument{%
  \providecommand\BibTeX{{%
    \normalfont B\kern-0.5em{\scshape i\kern-0.25em b}\kern-0.8em\TeX}}}

\setcopyright{acmlicensed}
\acmJournal{TWEB}
\acmYear{2022} \acmVolume{1} \acmNumber{1} \acmArticle{1} \acmMonth{1} \acmPrice{15.00}\acmDOI{10.1145/3572403}

\begin{document}

\title{\RoSGAS: Adaptive Social Bot Detection with Reinforced Self-Supervised GNN Architecture Search}

\author{Yingguang Yang}
\email{dao@mail.ustc.edu.cn}
\orcid{0000-0002-2473-6229}
\affiliation{%
  \department{School of Cyber Science and Technology}
  \institution{University of Science and Technology of China}
  \streetaddress{96 Jinzhai Road}
  \city{Hefei}
  \state{Anhui}
  \country{China}
  \postcode{230026}
}
\affiliation{%
  \institution{Key Laboratory of Cyberculture Content Cognition and Detection, Ministry of Culture and Tourism}
  \country{China}
}
\author{Renyu Yang}
\affiliation{%
  \institution{University of Leeds}
  \streetaddress{Woodhouse Lane}
  \city{Leeds}
  \state{West Yorkshire}
  \country{United Kingdom}
  \postcode{LS2 9JT}}
\orcid{0000-0001-6334-4925}
\authornote{Co-first author with equal contribution}
\email{r.yang1@leeds.ac.uk}
\author{Yangyang Li}
\affiliation{%
 \institution{National Engineering Laboratory for Public Safety Risk Perception and Control by Big Data}
 \streetaddress{11 Shuangyuan Road}
 \city{Beijing}
 \state{Beijing}
 \country{China}}
 \email{liyangyang@cetc.com.cn}
 \authornote{Corresponding author}
 \affiliation{%
  \institution{Key Laboratory of Cyberculture Content Cognition and Detection, Ministry of Culture and Tourism}
  \country{China}
}
\author{Kai Cui}
\affiliation{%
  \department{School of Cyber Science and Technology}
  \institution{University of Science and Technology of China}
  \streetaddress{96 Jinzhai Road}
  \city{Hefei}
  \state{Anhui}
  \country{China}
}
\affiliation{%
  \institution{Key Laboratory of Cyberculture Content Cognition and Detection, Ministry of Culture and Tourism}
  \country{China}
}
\email{kaicui@mail.ustc.edu.cn}
\author{Zhiqin Yang}
\author{Yue Wang}
\affiliation{%
  \institution{Beihang University}
  \streetaddress{37 Xueyuan Road}
  \city{Beijing}
  \state{Beijing}
  \country{China}
  \postcode{100191}}
\email{yangzqccc@buaa.edu.cn}
\email{zb2039111@buaa.edu.cn}
\author{Jie Xu}
\affiliation{%
  \institution{University of Leeds}
  \streetaddress{Woodhouse Lane}
  \city{Leeds}
  \state{West Yorkshire}
  \country{United Kingdom}
  \postcode{LS2 9JT}}
  \affiliation{%
  \institution{Beihang University}
  \streetaddress{37 Xueyuan Road}
  \city{Beijing}
  \country{China}}
\email{j.xu@leeds.ac.uk}
\author{Haiyong Xie}
\affiliation{%
  \department{School of Cyber Science and Technology}
  \institution{University of Science and Technology of China}
  \streetaddress{96 Jinzhai Road}
  \city{Hefei}
  \state{Anhui}
  \country{China}}
\affiliation{%
  \institution{Key Laboratory of Cyberculture Content Cognition and Detection, Ministry of Culture and Tourism}
  \country{China}
}
\email{hxie@ustc.edu.cn}
\authornote{Corresponding author}

\renewcommand{\shortauthors}{Yang, et al.}
\newcommand{\RoSGAS}{\textsc{RoSGAS}\xspace}

\begin{abstract}
\rule[0em]{13.95cm}{0.05em}
Social bots are referred to as the automated accounts on social networks that make attempts to behave like human. While Graph Neural Networks (GNNs) has been massively applied to the field of social bot detection, a huge amount of domain expertise and prior knowledge is heavily engaged in the state-of-the art approaches to design a dedicated neural network architecture for a specific classification task. Involving oversized nodes and network layers in the model design, however, usually causes the over-smoothing problem and the lack of embedding discrimination.
In this paper, we propose \RoSGAS, a novel \underline{R}einf\underline{o}rced and \underline{S}elf-supervised \underline{G}NN \underline{A}rchitecture \underline{S}earch framework to adaptively pinpoint the most suitable multi-hop neighborhood and the number of layers in the GNN architecture. More specifically, we consider the social bot detection problem as a user-centric subgraph embedding and classification task.
We exploit heterogeneous information network to present the user connectivity by leveraging account metadata, relationships, behavioral features and content features.
\RoSGAS uses a multi-agent deep reinforcement learning (RL) mechanism for navigating the search of optimal neighborhood and network layers to learn individually the subgraph embedding for each target user. A nearest neighbor mechanism is developed for accelerating the RL training process, and \RoSGAS can learn more discriminative subgraph embedding with the aid of self-supervised learning. Experiments on 5 Twitter datasets show that \RoSGAS outperforms the state-of-the-art approaches in terms of accuracy, training efficiency and stability, and has better generalization when handling unseen samples. 
\end{abstract}


\maketitle

\section{Introduction}

Social bots -- the accounts that are controlled by automated software and mock human behaviours \cite{abokhodair2015dissecting} -- widely exist on online social platforms such as Twitter, Facebook, Instagram, Weibo, etc. and normally have malicious attempts. For example, interest groups or individuals can use social bots to influence the politics and economy, e.g., swaying public opinions at scale, through disseminating disinformation and on-purpose propaganda, and to steal personal privacy through malicious websites or phishing messages \cite{varol2017online}. Such deception and fraud can reach out to a huge community and lead to cascading and devastating consequences.

Social bots have been long studied but not yet well-resolved due to the fast bot evolution \cite{cresci2020decade}. The third generation of bots since 2016 with deepened mixture of human operations and automated bot behaviors managed to disguise themselves and survived platform-level detection using traditional classifiers \cite{cresci2020decade,cresci2017paradigm}. The cat-and-mouse game continues -- while new work-around, camouflage and adversarial techniques evolve to maintain threats and escape from perception, a huge body of detection approaches emerge to differentiate the hidden behaviors of the emerging socials bots from legitimate users. The recent advancements in Graph Neural Networks (GNNs) \cite{scarselli2008graph} can help to better understand the implicit relationships between abnormal and legitimate users and thus improve the detection efficacy. GNN-based approaches \cite{liu2018heterogeneous,wang2019fdgars,breuer2020friend,feng2021botrgcn,feng2021heterogeneity,dou2020enhancing,peng2021reinforced} formulate the detection procedure as a node or graph classification problem. Heterogeneous graphs are constructed by extracting the accounts' metadata and content information from social networks before calculating numerical embedding for nodes and graphs. However, there are still several interrelated problems to be addressed:

\textit{GNN architecture design has a strong dependence upon domain knowledge and manual intervention.} In most of the existing works, the embedding results are inherently ﬂat because the neighbor aggregation ignores the difference between the graph structure pertaining to the target node and the structures of other nodes. This will result in the lack of deterministic discrimination among the final embeddings when the scale of the formed graph structure grows to be tremendous. To address this issue, subgraphs are leveraged to explore the local substructures that merely involve partial nodes, which can obtain non-trivial characteristics and patterns \cite{ullmann1976algorithm}.
However, the subgraph neural network based approaches heavily rely on experiences or domain knowledge in the design of rules for extracting subgraphs and of model architectures for message aggregation~\cite{lee2019graph,peng2020motif,alsentzer2020subgraph}. 
This manual intervention substantially impedes the elaborated design of a neural network model that can adapt to the evolving changes of the newer social bots. Using fixed 
and fine-grained subgraph extraction rules is not sufficiently effective \cite{ferrara2016rise,cresci2017paradigm}. 

\textit{Over-assimilated embedding when aggregating a huge number of neighbors.} The most intractable and demanding task is to effectively perceive and pinpoint the camouflages of the new-generation social bots. Camouflage technology mainly comprises two distinct categories -- feature camouflage and relation camouflage. Feature camouflage is referred to as the procedure where bots steal the metadata of the benign accounts and transform into their own metadata. They also employ advanced generation technology to create content information \cite{cresci2020decade}. Apart from mimicking features of legitimate users, relation camouflage techniques further hide the malicious behaviors by indiscriminately interacting with legitimate users and establishing friendships with active benign users \cite{yang2020rumor}. The interactions, particularly with the influencers, can considerably shield the bots from being detected. It is thus critical to include sufficient heterogeneous nodes in the neighborhood when extracting the subgraph for the target user so that camouflaged bots can be picked up.  Meanwhile, it is also important to differentiate the subgraph embeddings of different target users while similarizing the embeddings within the same target user.  However, the over-smoothing representation problem will manifest  \cite{li2018deeper,xu2018representation} as the involvement of a huge number of nodes in the GNNs tends to over-assimilate the node numerical embedding when aggregating its neighbor information.  


\textit{Inadequate labeled samples}. A large number of labeled samples are presumably acquired and massively used in the supervised model training. However, this assumption can be hardly achieved in the real-world social bot detection. In fact, there are always very limited users with annotated labels, or limited access to adequate and labelled samples \cite{gilani2017classification}. This will hamper the precision of supervised deep learning models and particularly lead to poor performance in identifying out-of-sample entities, i.e., the new types of social bots out of the existing datasets or established models. 

To address these issues, the state-of-the-art works \cite{xiong2017deeppath,zhou2019auto,zhong2020reinforcement,gao2020graph,lai2020policy} adopt reinforcement learning (RL) to search the GNN architecture.  However, such  approaches sometimes lack generalizability; the effectiveness of determining the optimal GNN architecture is tightly bound to specific datasets and usually have huge search space, and hence low efficiency. They are not suited for social network networks where the graph structure follows a power-law distribution \cite{barabasi1999emergence,muchnik2013origins}. In this scenario, dense and sparse local graph structures co-exist and huge disparities manifest among different users. Additionally, as only taking a small fraction of the labelled users as the environment state, the existing RL agents can hardly learn the state space in an accurate manner and will lead to a slow convergence in the RL agents.  Hence, it is highly imperative to personalize the selection of subgraphs and GNN architecture of the model for each target user, without prior knowledge and manual rule extraction, and to devise automated and adaptive subgraph embedding to fit  the ever-changing bot detection requirements.

In this paper, we propose \RoSGAS, a subgraph-based scalable \underline{R}einf\underline{o}rced and \underline{S}elf-supervised \underline{G}NN \underline{A}rchitecture \underline{S}earch approach to adaptively extract the subgraph width and search the model architecture for better subgraph embedding, and to speed up the RL model convergence through self-supervised learning. Specifically, we use Heterogeneous Information Network (HIN) to model the entities and relationships in the social networks and use meta-schema and meta-path to define the required relationship and type constraints of nodes and edges in the HIN, on the basis of real-world observations in social network platforms.
We formulate the social bot detection problem as a subgraph classification problem.
We firstly propose a multi-agent reinforcement learning (RL) mechanism for improving the subgraph embedding for target users. The RL agent can start to learn the local structure of the initial 1-order neighbor subgraph of a given target user and help to select the most appropriate number of neighbor hops as the optimal width of the subgraph. The RL agent is also elaborately devised to select the optimal number of model layers such that the neural networks are well-suited for encoding the dedicated subgraphs with sufficient precision, without introducing oversized architecture and computation overhead. We then exploit the residual structure to retain the characteristics of a target user as much as possible, thereby overcoming the over-smooth problem on the occasion of message aggregations from a huge number of neighbor nodes.
While using RL to automate the neighbor selection and model construction,  we additionally develop a nearest neighbor mechanism in the early stage of RL for accelerating the training process of the proposed RL approach. A self-supervised learning mechanism is further investigated and integrated with the \RoSGAS framework to overcome the deficiency of over-assimilated embedding. The self-supervised module can facilitate more discriminative representation vectors and help to enhance the  capability of expressing discrepancies among different target users.
Experimental results show that \RoSGAS outperforms the state-of-the-art approaches over five Twitter datasets and can achieve competitive effectiveness when compared with hand-made design of the model architecture.  

Particularly, the main contribution of this work are summarized as follows:
\begin{itemize}
\item proposed for the first time a user-centric social bot detection framework based on Heterogeneous Information Network and GNN without prior knowledge.
\item developed an adaptive framework that leverages Deep Q-learning to optimize the width of the subgraph and the GNN architecture layers for the subgraph corresponding to each target user. 
\item investigated a nearest neighbor mechanism for accelerating the convergence of training process in RL agents. 
\item proposed a self-supervised learning approach to enable homologous subgraphs have closer representation vectors whilst increasing the disparities of representation vectors among non-homologous subgraphs after information aggregation.
\item presented an explicit explanation for the model stability.
\end{itemize}

\textit{Organization}. This paper is structured as follows: Section \ref{sec:problem} outlines the problem formulation, and Section \ref{sec:method} describes the technical details involved in \RoSGAS.
The experimental setup is described in Section \ref{sec:exp_setup} and the results of the experiment are discussed in Section \ref{sec:exp_res}. More discussions are given in Section \ref{sec:discussion}. Section \ref{sec:related} presents the related work before we conclude the paper in Section \ref{sec:conclusion}.
\begin{figure}[t]
\centering
\includegraphics*[width=0.68\textwidth]{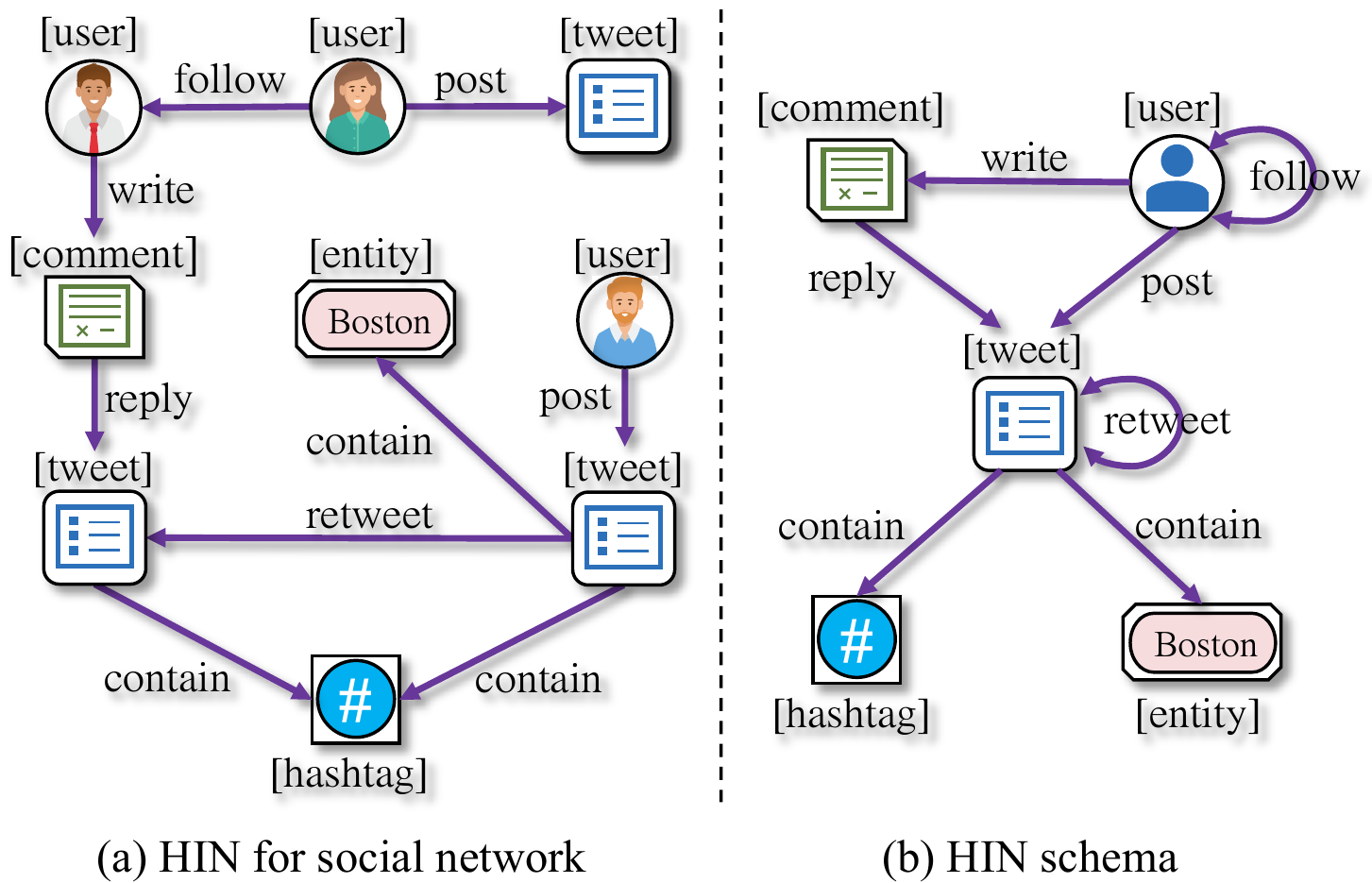}
\caption{(a) An example of HIN for social network. (b) Network schema of the HIN for social network.\label{fig1}}
\label{fig:graph}
\end{figure}
\begin{figure}[t]
\centering
\includegraphics*[width=0.6\textwidth]{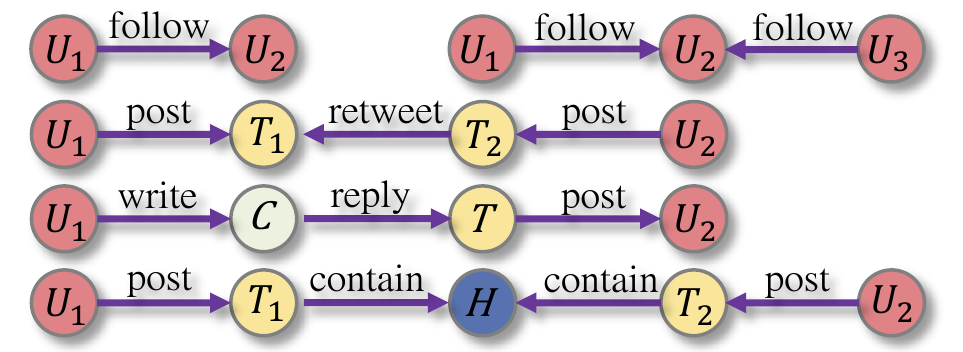}
\caption{The extracted meta-paths.}
\label{fig:meta-path}
\end{figure}

\section{Problem formulation} 
\label{sec:problem}

In this section, we introduce HINs and information network representation before discussing the scope of this work and formulating the target problem.

\subsection{Preliminaries}
\label{sec:problem:preliminaries}

In this work, we follow the terminologies used in the work of ~\cite{dong2017metapath2vec,shi2016survey,peng2021lime,hei2021hawk} to define Heterogeneous Information Network (HIN) embedding. The aim is to project different nodes in the HIN into a low-dimensional space whist preserving the heterogeneous structure and relationships between heterogeneous nodes.

\noindent
\textbf{Definition 1. Heterogeneous Information Network.}
\label{df:HIN}
A heterogeneous information network (HIN) denoted as $\mathcal{G}=$ $\mathcal{G}(\mathcal{V},\mathcal{E},\mathcal{F},\mathcal{R},\varphi,\phi)$, where $\mathcal{V}$ denotes the nodes set, $\mathcal{E}$ denotes the edges set, $\mathcal{F}$ denotes the node types set and $\mathcal{R}$ denotes the edge types set. In real-world settings, there may be multiple types of nodes or edges, i.e., $|\mathcal{F}|+|\mathcal{R}|>2$.
Each individual node $i\in\mathcal{V}$ is associated with a node type mapping function $\varphi:\mathcal{V}\rightarrow\mathcal{F}$; similarly, each individual edge $e\in\mathcal{E}$ has an edge type mapping function $\phi:\mathcal{E}\rightarrow\mathcal{R}$.

In a nutshell, real-life information networks have different structures consisting of multi-typed entities and relationships. A relationship is referred to as the link between entities in a network or graph system. For example, Fig.~\ref{fig1}(a) shows an example of social network HIN that we construct for Twitter. It comprises five types of nodes (user, tweet, comment, entity and hashtag) and six types of relationships (write, follow, post, reply, retweet, and contain).

\noindent
\textbf{Definition 2. Network Schema.}
\label{Network_schema}
Given a HIN $\mathcal{G}(\mathcal{V},\mathcal{E},\mathcal{F},\mathcal{R},\varphi,\phi)$, the network schema for network $\mathcal{G}$ can be denoted as $\mathcal{T}(\mathcal{F},\mathcal{R})$, a directed graph with the node type set $\mathcal{F}$ and edge types set $\mathcal{R}$. In simple words, HIN schema comprehensively depicts the node types and their relations in an HIN, and provide a meta template to guide the exploration of node relationships and extract subgraphs 
from the HIN. Fig.~\ref{fig1}(b) exemplifies the network schema that can reflect entities and their interactions in a generic social network.

\noindent
\textbf{Definition 3. Meta-path.} 
Given a Meta Schema $\mathcal{T}(\mathcal{F},\mathcal{R})$, a Meta-Path $\mathcal{MP}$, denoted as $\mathcal{F}_1\stackrel{\mathcal{R}_1}{\longrightarrow}\mathcal{F}_2\stackrel{\mathcal{R}_2}{\longrightarrow}\ldots\stackrel{\mathcal{R}_{l-1}}{\longrightarrow}\mathcal{F}_l$, is a path on $\mathcal{T}$ that  connects  a pair  of  network  nodes and defines the composite relation $\mathcal{R}$ which contains multi types of relationships.

In reality, a meta-path describes the semantic relationship between nodes. Mining such semantic relationship is the cornerstone of subsequent tasks such as classification, clustering, etc. As shown in Fig.~\ref{fig:meta-path}, we extracted five most useful meta-paths from our defined meta-schema, based on observations in social network platforms. 

\subsection{Problem Statement}
We consider the social bot detection problem as a subgraph classification task, instead of a node classification task, in a semi-supervised learning manner.

\noindent
\textbf{Definition 4. Semi-supervised Subgraph Classification.} \label{def:subgraph_classification} 
Given a collection of target users, the subgraph pertaining to the $i$-th target user can be defined as $\mathcal{G}_i=\{\mathcal{V}_i,\mathcal{X}_i,\{\mathcal{E}_r\}|^R_{r=1},y_i\}$. $\mathcal{V}_i$ is the collection of nodes in the subgraph, including $v_{i_0}$, the target user itself, and the neighbors $\{v_{i_1},\ldots,v_{i_n}\}$ within a few hops from the target user.
These nodes are extracted from the entire graph $\mathcal{G}$ and consist of different types. 
Each node initially has a $d$-dimensional feature vector and $\mathcal{X}_i$ represents the vector set of all node embeddings, i.e.,
$\mathcal{X}_i=\{x_{i_0},\ldots,x_{i_n}\}$ where each element
$x\in\mathbb{R}$. 
The edge in the subgraph can be represented as $e^r_{i_m,i_n}=(v_{i_m},v_{i_n})\in\mathcal{E}_r$, where $v_{i_m}$ and $v_{i_n}$ is connected through a certain relationship $r\in\{1,\ldots,R\}$.
$y_i\in\{0,1\}$ represents a binary label of the target user $v_{i_0}$; 0 indicates benign account while 1 represents a social bot.
Once a dedicated subgraph $G_i$ is extracted, the subsequent task is to conduct a subgraph binary classification. At the core of the adaptive architecture search is to pinpoint the subgraph width ($k$) and 
the optimal value of model layers ($l$) that constitute the whole bot detection model.

\subsection{Research Questions}

For achieving discriminative, cost-effective and explainable subgraph embedding, there are three main research challenges facing the RL-based social bot detection:

\textbf{[Q1] How to determine the right size of the subgraph for an individual target user in a personalized and cost-effective manner?} The major issue with the DNN construction is the selection of neighbor hops and the model layers. In fact, two interrelated yet opposite factors may affect the choice of a detection model. On the one hand, a larger number of hops and model layers can involve more nodes, including both benign and malicious nodes, in the neighbor aggregation. This is beneficial for the detection quality since the hidden camouflages of social bots could be identified by the higher-order semantic embedding enabled by the continuum of HIN-based data engineering and DNN model training. However, excessive involvement will bring performance issues in terms of time- and computation- efficiency, and, more severely, lead to the over-smooth problem commonly manifested in graphs at scale \cite{li2018deeper}. On the other hand, a small portion of the neighbors would overlook node information and lead to less informative node embeddings. To resolve this dilemma -- balancing competitive accuracy and high computation efficiency -- whilst addressing the assimilation within the neighborhood, it is critical to automatically pick up an appropriate hops of neighbors and to stack up \textit{just-enough} neural network layers to be assembled in the detection model. This requires the reinforcement learning process to properly define dedicated policies and optimize the setting of environment states and actions.


\textbf{[Q2] How to accelerate the convergence of reinforcement learning?}  It is observable that in the initial stage of training, the learning curve substantially fluctuates and this phenomenon will slow down the training process and model convergence. This is because in the starting phase the noisy data may take up a high portion of the limited memory buffer and thus misdirect to the wrong optimization objective.  To ensure the training efficiency, it is necessary to boost the action exploration in RL agent and speed up the training stabilization. 

\textbf{[Q3] How to more efficiently optimize the reinforcement learning agents in the face of limited annotated labels?} Data annotation is expensive and sometimes difficult in practical problem solving. If only a small portion of the labelled users are used as the input of the RL agent as environment state, the state space cannot be accurately and efficiently learnt within a required time frame. This will consequently delay the optimization of a RL agent and further have a cascading impact on the multi-agent training. This issue therefore necessitates an self-supervised learning mechanism for optimizing the training effectiveness and efficiency.


\section{methodology}
\label{sec:method}

In this section, we will introduce how we design the social bot detection framework through adaptive GNN architecture search with reinforcement learning. 
We first introduce the basic process of subgraph embedding (Section \ref{sec:method:subgraph_embedding}). In response to \textbf{[Q1]}, we introduce a reinforcement learning enhanced architecture search mechanism (Section \ref{sec:method:reinforced}). To address \textbf{[Q2]}, we propose a nearest neighbor mechanism (Section \ref{sec:method:nearest}) for accelerating the convergence process of reinforcement learning. The self-supervised learning approach is discussed (Section \ref{sec:method:self_supervised}) to tackle \textbf{[Q3]} before we present how to tackle parameter explosion and outline the holistic algorithm workflow  (Section \ref{sec:method:parameter} and Section \ref{sec:method:algorithm}).

\begin{table}[t]
\centering
\caption{Notations.}
\label{tab:notation}
\scalebox{0.95}{
\begin{tabular}{r|l}
\hline
\toprule
\textbf{Symbol}       & \textbf{Definition}  \\ 
\midrule
$U_i$; $T_i$; $C_i$; $H_i$  & User; tweet; comment; hashtag   \\
$\mathcal{G}; \mathcal{G}_i; \mathcal{V}; \mathcal{E}$ & Graph; $i$-th subgraph; node set; edge set \\
$\mathcal{F}; \mathcal{R}$ & Node type; relation type \\
$\varphi; \phi$ & Mapping a node to the type $\mathcal{F}$; mapping a edge to the type $\mathcal{R}$ \\
$\mathcal{V}_i; \mathcal{X}_i$ & Node set of subgraph $\mathcal{G}_i$; node feature set of subgraph $\mathcal{G}_i$ \\
$v_{i_0}; v_{i_n}; y_i$ & $i$-th target user; $n$-th node in Subgraph $\mathcal{G}_i$; label of $i$-th target user \\
$e^r_{v_{i_m},v_{i_n}}$; $r$ & An edge connecting $v_{i_m},v_{i_n}$ through a relationship $r\in\mathcal{R}$ \\
$k; l$ & The number of neighbor hops and the number of model layers \\
$\mathcal{D};h_j;m_i$ & The target user set; the original feature of $v_j$; the model for target user $v_{i_0}$ \\
$L;\rm{L}$ & The last layer number and the error correction parameter \\
$\alpha^k_{ij}; W^k$ & The $k$-th attention coefficient and the $k$-th weight matrix \\
$z_i;\pi$ & The embedding of the $i$-th subgraph and the policy network \\
$s_t;a_t;r_t$ & The state, chose action, and the reward at timestamp $t$  \\
$\mathcal{R},b$ & The advantage function measurement of $(s_t,a_t)$ and the history window size \\
$Q,\gamma$ & The value of state-action pair and the future cumulative rewards discount parameter \\
$B;\tau$ & The observation experience set and the state-action pair \\
$d_\tau,\Theta$ & The state distance measurement function and the parameter of RL agent \\
$\alpha_0,\beta$ & The initial weight and decay rate of $\alpha$ in the nearest neighbor mechanism \\
$\lambda$ & The weight parameter of the GNN loss function \\
$\mathcal{G}_i^k$ & A subgraph $\mathcal{G}$ for  $i$-th target user with $k$ hop neighbors \\
${\mathcal{L}_{pretext}}_i,g_i$ & The loss of the $i$-th pretext task, the $i$-th stacked GNN encoder  \\
${y_{pretext}}_i$ & the ${y_{pretext}}_i$ stands for the ground truth of subgraph $\mathcal{G}_i$ acquired by pretext task\\
$\mathcal{G}_i'$, $\bar{\mathcal{G}_i}$ & $\mathcal{G}_i'$ is a positive sample for $\mathcal{G}_i$. $\bar{\mathcal{G}_i}$ is a negative sample for $\mathcal{G}_i$. \\
$z_i', \bar{z_i}$ & $z_i'$ is the representation of $\mathcal{G}_i'$, $\bar{z_i}$ is the representation of $\bar{\mathcal{G}_i}$\\

\bottomrule
\end{tabular}}
\end{table}

\subsection{Overview}
\label{sec:method:subgraph_embedding}

\begin{figure}[t]
  \centering
  \includegraphics[width=\linewidth]{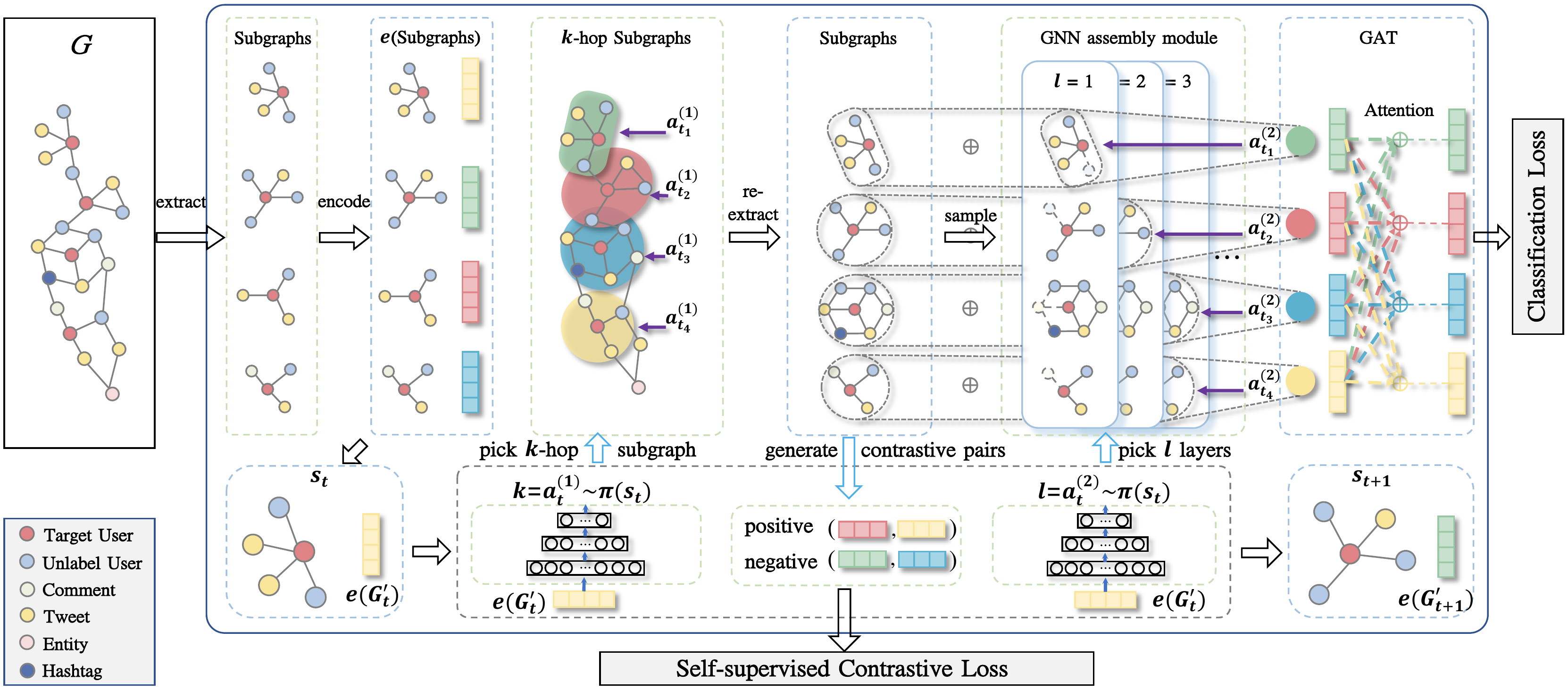}
  \caption{The proposed \RoSGAS framework.}
  \label{fig:framework}
\end{figure}

Fig.~\ref{fig:framework} depicts the overall architecture of \RoSGAS to perform the subgraph embedding. The workflow mainly consists of three parts: graph preprocessing and construction, RL-based subgraph embedding and the final-stage attention aggregation mechanism among graph nodes before feeding into the final classifier for determining the existence of social bots. To aid discussion, Table \ref{tab:notation} depicts the notations used throughout the paper.

\subsubsection{Graph Construction}

Initially, the feature extraction module transforms the original information into a heterogeneous graph. The edges between nodes in the heterogeneous graph are established based on the account's friend relationships and interactions in the social network platform. We retrieve the meta features and description features of each account as its initial node feature in a similar way as \cite{yang2020scalable}. Extra tweet features and entities are extracted by using NLPtool\footnote{\url{https://github.com/explosion/spaCy}} from the original tweets. The composition of features for each type of node may vary. For \textit{user} node, features such as \textit{status}, \textit{favorites}, \textit{list}, etc. are extracted from user metadata. For \textit{tweet} node, we normally extract the number of retweets and the number of replies, whilst embedding the tweet content into a 300-dimension vector and concatenating them together as their original features. Similarly, we embed the content of \textit{hashtag} and \textit{entity} to 300-dimension vectors. To simplify the feature extraction and processing, the feature vector of each node type is set to be 326 dimensions. Those with insufficient dimensions are filled with zero. More details will be given in Section~\ref{sec:exp_setup:feature_extraction}.

To refine the heterogeneous graph under the given semantics, we further conduct a graph pre-processing by enforcing meta-paths upon the original graph and only the entities and edges conforming the given meta paths will be retained in the graph structure. As shown in Fig.~\ref{fig:meta-path}, we extracted five meta-paths that are widely-recognized in social graphs and represent most of the typical behaviors in the meta-schema defined in Section \ref{Network_schema}. The main purpose is to cut down the information redundancies in the graph at scale and thus substantially improve the computational efficiency. The transformed graph will be further used to extract subgraphs for the target users before feeding the subgraphs into the down-streaming tasks including the numerical embedding and classification. 



\subsubsection{RL-based Parameter Searching}

The primary goal of the parameter selection is to determine the appropriate subgraph width ($k$) and the model layers ($l$) for each target user in the target user collection $\mathcal{D}$ with $n$ users.
In a nutshell, for each target user, we take it as the center node and first extract a fixed width (e.g., 1 hop) subgraph $\mathcal{G}_i$ as the initial subgraph.  Afterwards, we encode the subgraph into the embedding space and regard the embedding as the environment state before feeding it into the reinforcement learning agent. To be more specific, the embedding representation of the $i$-th target user can be obtained by using an encoder (e.g., average encoding operation, sum encoding operation, etc.):
\begin{equation}
    e(\mathcal{G}_i) = \{h_j | v_j \in \mathcal{V}_i(\mathcal{G}_i)\}.
\label{eq2}
\end{equation}

At the core of the embedding improvement is the RL agent. The preliminary encoding result will be fed into the policy $\widetilde{\pi}_1$ and  $\widetilde{\pi}_2$ in the RL agent, successively.
$\widetilde{\pi}_1$ is responsible for selecting the appropriate width of subgraph $\mathcal{G}_i'$ while $\widetilde{\pi}_2$ is in charge of pinpointing the most suitable number of layers for constructing model $m_i$ for the target user $v_{i_0}$. The types of specific model layer can be selected from the most popular models such as GCN~\cite{GCN}, GAT~\cite{GAT} and GraphSAGE~\cite{GraphSAGE}, etc.
Generally speaking, the goal of reinforcement learning is to maximize the expected accuracy:  $\mathbb{E}[\mathcal{R}_\mathcal{D}(\{m_i,\mathcal{G}_i\}|^n_{i=1})]$ on $\mathcal{D}$: 
\begin{equation}
    \widetilde{\pi}^*_1,\widetilde{\pi}^*_2 = \operatorname*{argmax}_{\theta_1,\theta_2}\mathbb{E}
        [\mathcal{R}_\mathcal{D}
            (\{\pi_1(e(\mathcal{G}_i);\theta_1), \pi_2(e(\mathcal{G}_i);\theta_2)\}|^{n}_{i=1})].
    \label{eq:eq1}
\end{equation}
More details about the learning procedure will be discussed in Section \ref{sec:method:reinforced}.


To calibrate the subgraph embeddings, we can then aggregate the $k$-hop neighbors and  stack models with various $l$ layers.
However, the aggregation from the stacked GNN models would blur the original detectable features of the target user in the subgraph embedding.
To mitigate this issue, we apply the residual network to aggregate the target node's input features and its corresponding embedding  delivered by the last layer of the model:

\begin{equation}
    h_j^{(L)} = ADD(x{_{i_j}},h_j^{(L)}),  \label{eq3}
\end{equation}
where $L$ is the last layer of the stacked GNN model. Then we can apply a pooling operation (e.g., average, sum, etc.) to integrate the subgraph $\mathcal{G}_i$ into a super node:
\begin{equation}
    z_i^{(L)} = READOUT(\{h_j^{(L)}\}^{n'}_{j=1}).  \label{eq4}
\end{equation}

\subsubsection{Attention Aggregation and Classification}

We adopt an attention mechanism for integrating the influence of subgraphs belonging to the relevant neighbors into the final embedding:
\begin{equation}
    z_i = \frac{1}{K}\sum\limits_{k=1}^K\sum\limits_{\mathcal{G}_j\in\mathcal{G}} \alpha^k_{ij}{\rm{W}^k}z_j^{(L)}, \label{eq5}
\end{equation}
where $\alpha_{ij}$ is the attention coefficient, $\rm{W}\in\mathbb{R}^{d_L\times{d_l}}$ is the weight matrix and $K$ is the number of independent attention. 
$z_i$ is the final embedding for detecting if the target user is a social bot.
Eventually, the bot classifier digests the learned vector embeddings to learn a classification model and determines if a given social user behaviors normally or maliciously. General purpose techniques including Random Forest, Logistic Regression, SVM, etc. can be adopted for implementing the classifier.

\subsection{Reinforced Searching Mechanism}
\label{sec:method:reinforced}

In this subsection, we will introduce how to obtain the optimal policies $\tilde{\pi}^*_1$ and $\tilde{\pi}^*_2$ through the searching mechanism. The learning procedure of the optimal $\tilde{\pi}^*_1$ and $\tilde{\pi}^*_2$ can be formulated as a Markov Decision Process (MDP). An RL agent episodically interacts with the environment where each episode lasts for $T$ steps. The MDP includes state space, action space, reward and the transition probability that maps the current state and action into the next state.  

\mypara{State Space.} In each timestamp $t$, the state $s_t$ is defined as the embedding of the subgraph extracted from $\mathcal{G}$.

\mypara{Action Space.} Since we need two policies to pinpoint the optimal width of subgraph and the optimal number of model layers respectively, the action at timestep $t$ consists of dual sub-actions $(a^{(1)}_t,a^{(2)}_t)$. The RL agent integrated in our proposed framework \RoSGAS performs an action $a^{(1)}_t$ to get the value of $k$, and performs an action $a^{(2)}_t$ to get the value of $l$.
For instance, $a^{(1)}_t$ is chosen at the timestep $t$ to re-extract the subgraph of the target user $v_{i_0}$. We then calculate the number of reachable paths from target user $v_{i_0}$ to the other target users in this subgraph as the connection strength. For those target users that are included in the collection $\mathcal{D}$ yet excluded from this subgraphs, the connection value will be set to $0$. $L1$ normalization is performed upon these values as the reachability probabilities from the target user to the other target users. 
After selecting certain actions $(a^{(1)}_t,a^{(2)}_t)$ at the timestep $t$, the RL environment forms a probability distribution $P_i$.

\mypara{Transition.}
The probability $P_i$ serves as the state transition probability of the reinforcement learning environment. The subgraph embedding of any other target user is used as the next state $s_{t+1}$. The whole trajectory of the proposed MDP can be described as $(s_0, (a^{(1)}_0,a^{(2)}_0 ), r_0, \ldots, s_{T-1}, (a^{(1)}_{T-1}, \\a^{(2)}_{T-1} ), r_{T-1}, s_T)$.

\mypara{Reward.}
We need to evaluate whether the search mechanism is good enough at the current timestep $t$. In other words, it reflects if the parameters in the current RL agent can achieve better accuracy than the parameters at the previous timestep $t-1$. To do so, we firstly define a measure to flag the improvement of model accuracy when compared with the previous timesteps:
\begin{equation}
    \mathcal{R}\left(s_t,a_t\right)=(\mathcal{ACC}\left(s_t,a_t\right)-  \frac{\sum_{i=t-b}^{t-1} \mathcal{ACC} \left(s_i,a_i\right)}{b-1}),\label{eq6}
\end{equation}
where $b$ is a hyperparameter that indicates the window size of historical results involved in the comparison and $\mathcal{ACC}\left(s_i,a_i\right)$ is the accuracy of subgraph classification on the validation set at the timestep $i$.
$\frac{\sum_{i=t-b}^{t-1} \mathcal{ACC} \left(s_i,a_i\right)}{b-1}$ reflects the average accuracy in the most recent $b$ timestep windows.
The training RL agent continuously optimizes the parameters to enable a rising accuracy and accordingly positive rewards. This will give rise to the cumulative rewards in finite timesteps and eventually achieve the optimal policies.

We use a binary reward $r_t$ combined with Eq.\ref{eq6} to navigate the training direction as follows:
\begin{equation}
r(s_t,a_t)=\left\{
             \begin{array}{lr}
             1, & \rm{if}\ \mathcal{R}\left(s_{t},a_{t}\right) > \mathcal{R}\left(s_{t-1},a_{t-1}\right) \\
             -1, & \rm{otherwise}.  
             \end{array}
\right.   
\label{eq7}
\end{equation}
The value is set to be 1 if $a_t$ can increase the $\mathcal{R}$ compared with that of the previous timestep $t-1$; otherwise it will be set -1.

\mypara{Termination.}  State-action values can be approximated by the Bellman optimal equation:
\begin{equation}
    Q^*{\left(s_t,a_t\right)} = r\left(s_t,a_t\right)+ \gamma\ \mathop{\rm arg\  max}\limits_{a^{'}}Q^*(s_{t+1},a^{'}). \label{eq7}
\end{equation}
Nevertheless, to improve both training speed and stability, we will introduce an enhanced approximation approach in Section \ref{sec:method:nearest}.   We exploit the $\epsilon$-greedy policy to select the action $a_t$ with respect to $Q^*$ and obtain the policy function $\pi$:
\begin{equation}
    \pi\left(a_t|s_t;Q^*\right)=\left\{
             \begin{aligned}
             {\rm random\ action},\quad &  w.p. \epsilon \\
             \mathop{\rm argmax}\limits_{a}Q^*\left(s_t,a\right),\quad & \rm{otherwise}.  
             \end{aligned}\label{eq8}
\right.
\end{equation}

\subsection{Nearest Neighbor Mechanism for Accelerating Model Stabilization}
\label{sec:method:nearest}

Conventionally, at each time step $t$, the RL agent employs its prediction network to determine the value of state-action pairs for choosing the best action to maximize the future cumulative rewards. Inspired by \cite{shen2021theoretically}, we applied the nearest neighbor mechanism for assisting and accelerating the training process of the RL agent. The intuition behind the scheme is that when the RL agent observes similar or the same state-action pairs, the environment is highly likely to produce a similar reward value. In other words, the distance between state-action pairs can indicate their relative reward values.
Therefore, we aim to find out the similar state-action pairs and determine the reward of the current state-action pair by combining the reward estimated by the Q-network (i.e., the prediction network) and the reward of the existing similar pairs. This means that the model training benefits from both the RL environment and the prediction network, which can boost the action exploration and accelerate the training stabilization. 


To look into and record the historical actions, we set up an observation experience set $B=\{\tau_1,\ldots,\tau_n\}$; each element $\tau_i$ in $B$ represents a pair of explored state $s_i$ and the corresponding selected action $a_i$, namely, $(s_i,a_i)$.
While recording the action-state pairs, we also record the corresponding value labels $\{Q(\tau_i)\}\subseteq\mathbb{R}$.
We employ the distance function $d_\tau$ -- for example using  cosine to calculate the similarity -- to measure the distance between the explored state-action pairs and the incoming state-action pairs.
We use the distance to ascertain the nearest neighbor of the state-action pair to be estimated from the recorded state-action pairs.
Subsequently, the value label of the nearest neighbor can be used to estimate the value of the state-action pair:
\begin{equation}
\hat{Q}(\tau) = \rm{min}\{Q(\tau_i)+L\cdot d_\tau(\tau,\tau_i)\} |^{n}_{i=1},  \label{eq:10}
\end{equation}
where $\hat{Q}(\tau)$ is the estimated value of $\tau$, and $\rm{L}$ is a parameter to correct the estimated error.

We combine $\hat{Q}(\tau)$ estimated by the nearest neighbor mechanism and $Q(s_{t},a_{t};\Theta^{target})$ estimated by the target network into a new estimated value $\hat{Q}(s_{t},a_{t})$:
\begin{equation}
   \hat{Q}(s_{t},a_{t}) =  \alpha\cdot\hat{Q}(\tau_t) + (1-\alpha)\cdot(r+\gamma {\rm{max}}_{a_{t+1}}Q(s_{t+1},a_{t+1};\Theta^{target})),
\end{equation}
where $\alpha$ is an exponentially decaying weight parameter and $\hat{Q}(\tau_t)$ is the estimated value of $\tau_t=(s_t,a_t)$ by using the nearest neighbor search mechanism. 
In fact, $\alpha$ is used to assist the RL optimization in the early stage, and gradually reduce the effect of the proposed nearest neighbor mechanism when the training procedure moves forward. To achieve this, we set $\alpha=\alpha_0\cdot(1-\beta)^k$,  where $\alpha_0\in(0,1]$ is an initial weight; $\beta\in(0,1)$ is a decay rate and $k$ is the episode number. We then define the RL loss function as:
\begin{equation}
L(\Theta) = (\hat{Q}(s_{t},a_{t}) - Q(s_{t},a_{t};\Theta^{pred}))^2, \label{eq:11}
\end{equation}
where $\Theta^{pred}$ is the parameters of the DQN agent's prediction network, and the $\Theta^{target}$ is the parameters of the target network.

\subsection{Self-supervised Learning}
\label{sec:method:self_supervised}

To better differentiate the subgraph representations among graphs, we propose a contrastive self-supervised learning approach to maximize the difference between two distinct patches, without relying on the human-annotated data samples. The task of self-supervised learning (also known as pretext task) is to minimize the distance between positive samples whilst maximizing the distance between negative samples. In the context of this work, all substructures (e.g., subgraphs with 1-hop neighbors or 2-hop neighbors) pertaining to the same user should have similar representation vectors. Subgraphs belonging to two distinct target users should have discriminative embeddings.  

In general, given a subgraph $\mathcal{G}_i=\{\mathcal{V}_i,\mathcal{X}_i,\{\mathcal{E}_r\}|^R_{r=1},y_i\}$, 
the loss $\mathcal{L}_{pretext}$ for a self-supervised learning task can be defined as follows:
\begin{equation}
 \mathcal{L}_{pretext}(\mathcal{A}_i,\mathcal{X}_i,g_i)=\phi(g_i(\mathcal{G}_i),y_{pretext_i}),  \label{eq:12}
\end{equation}
where $g_i$ is the stacked GNN encoder for the extracted subgraph $\mathcal{G}_i$ and the $y_{pretext_i}$ stands for the ground truth of subgraph $\mathcal{G}_i$ acquired by a specific self-supervised pretext task.  

Practically, the key step is to generate positive samples in our self-supervised pretext task. After the RL algorithm outputs the customized width value $k$ for the target node $v_i$, we random select a $\bar{k}\in[1,K]!=k$ as the width of a new subgraph $\mathcal{G}_i'$, to serve a new positive sample of the original subgraph $\mathcal{G}_i$. Meanwhile, to provide the  negative-sampled counterparts, we randomly select the target user $v_j$ from the target user set $\mathcal{D}$ and directly use the learnt value $k$. We use the stacked GNN encoder $g_i$ and the proposed RL pipelines to perform feature extraction and summary, and obtain the final subgraph embedding $z_i$, $z_i'$, $\bar{z_i}$ for the subgraphs $\mathcal{G}_i$, $\mathcal{G}_i'$, $\bar{\mathcal{G}_i}$, respectively. Then we use the margin Triplet loss for model optimization to obtain high-quality representations that can well distinguish the positive and negative samples. The loss function is defined as follows:
\begin{equation}
\mathcal{L}_{pretext_i}=-max(\sigma(z_iz_i')-\sigma(z_i\bar{z_i})+\epsilon,0),  \label{eq:13}
\end{equation}
where $\epsilon$ is the margin value. The loss function of the pretext task will be incorporated into the holistic loss function as the optimization objective of \RoSGAS.
Since there may exist many overlapping nodes between different subgraphs, especially in a large-scale social graph, the adoption of the loss function can effectively avoid excessive differentiation between positive and negative samples and prevent from any performance degradation of representation.

\subsection{Parameter Sharing and Embedding Buffer Mechanism}
\label{sec:method:parameter}

The customized model construction for each individual target user will lead to a substantial number of training parameters. We use the following two schemes to alleviate this problem.

\begin{itemize}
    \item \textit{Parameter Sharing}: We first determine a maximum base layer number $k$ to initialize the model, and then repeatedly stack the whole or part of these layers according to $a_t$ output from the RL agent in each timestep $t$ in the initialization order. This can avoid training a large number of model parameters.
    \item \textit{Embedding Buffer}: We buffer the embeddings of the relevant subgraphs as a batch to carry out $a_t^{(2)}$ in each timestep to reduce unnecessary operations for embedding passing. Once the buffer space approaches the specified batch size, the model re-construction  will be triggered by leveraging the obtained number of layers from $a_t^{(2)}$ and adopting the buffered embeddings. We cleanse the buffer space once the GNN model training terminates to ensure the buffer can be refilled in the later stage.
\end{itemize}

\begin{algorithm}[t]
\caption{The overall process of \RoSGAS}\label{algorithm1}
\KwIn{The max neighbor hop number, layer number: $K$ and $L$; initial neighbor hop number $k_{init}$, the batch size of GNN and DQN: $B_{G}$, $B_{D}$, DQN training step $S$, the total training epoch $T$, epsilon value $\epsilon$, the window size $b$, the error correction parameters $\rm{L}$, the initial decay parameter $\alpha_0$, the decay rate $\beta$ ,the full graph $\mathcal{G}$, targeted node set $V$.}
Initialize ${L}$ GNN layers, RL agent networks ${\pi}_1$, ${\pi}_2$\; \label{algl:1}
Initialize the memory buffer $\mathcal{M}^{(1)}$, $\mathcal{M}^{(2)}$, and the GNN buffer $\mathcal{B}$ \; \label{algl:2}
Sample a target node, extract $k_{init}$-hop subgraph $\mathcal{G}_{0}$\; \label{algl:3}
$s_0=e(\mathcal{G}_0)$ via Eq. \ref{eq2}\; \label{algl:4}
\For{$t=0,1,\ldots,T$}
{
    $a_t=(a^{(1)}_t,a^{(2)}_t)$ via Eq.~\ref{eq8}\; \label{algl:6}
    Re-extract subgraph $\mathcal{G}^{'}_{t}$ and $e(\mathcal{G}^{'}_t)$ via Eq. \ref{eq2}\; \label{algl:7}
    Sample meta-path instances to get  new $\mathcal{G}^{'}_{t}$\; \label{algl:8}
    Store $\mathcal{G}^{'}_t$ and $a^{(2)}_t$ into buffer $\mathcal{B}$\; \label{algl:9}
    \If{${\rm size}(\mathcal{B})>B_{D}$}{ \label{algl:10}
        Stack $a^{(2)}_t$ layers GNN \; \label{algl:11}
        \For{$b=1,\ldots,B_G$}{ \label{algl:12}
            Generate positive and negative pair for the $b$-th target user in $\mathcal{B}$\; 
            Train the stacked model on the buffer of action $a^{(2)}_t$ via Eq. \ref{eq14}\; \label{algl:14}
        } \label{algl:14}
        Clear the buffer for $a^{(2)}_t$ in $\mathcal{B}$\; \label{algl:15}
    } \label{algl:16}
    Obtain $r_t$ on validation dataset via Eq. \ref{eq6}\; \label{algl:17}
    Jump to the next subgraph $\mathcal{G}_{t+1}$ and $e(\mathcal{G}_{t+1})$\; \label{algl:18}
    Store the $T^{(1)}_t=(s_t,a^{(1)}_t,s_{t+1},r_t)$ into $\mathcal{M}^{(1)}$\; \label{algl:19}
    Store the $T^{(2)}_t=(s_t,a^{(2)}_t,s_{t+1},r_t)$ into $\mathcal{M}^{(2)}$\; \label{algl:20}
    \For{{\rm step}=1,\ldots,S}{ \label{algl:21}
        Optimize ${\pi}_1$ and ${\pi}_2$ via Eq. \ref{eq:11} \; \label{algl:22}
    } \label{algl:23}
} \label{algl:24}
Re-init GNN to train via Eq. \ref{eq14} with $\widetilde{\pi}^*_1,\widetilde{\pi}^*_2$ \; \label{algl:25}
\end{algorithm}

\subsection{Put Them Together}
\label{sec:method:algorithm}


Algorithm ~\ref{algorithm1} summarizes the overall training process of the proposed \RoSGAS including the initialization of the subgraph embedding and the follow-up architecture search via the proposed RL process.
Specifically, We first construct the social graph according to our definition in Section \ref{df:HIN} before initializing the GNN model with the max layers $L$ and the parameters of the two RL agents (Line 1).
At the training stage, we randomly sample a target user and embed its $k_{init}$-hop subgraph as the initial state $s_0$ (Lines 2-4). Afterwards, at each time step, an action pair was chosen to indicate the width $k$ of the subgraph and the number $l$ of GNN layers for the target user (Line 6). Then we re-extract subgraph $\mathcal{G}^{'}_t$, store $\mathcal{G}^{'}_t$ and the value $l$ represented by $a^{(2)}_t$ into the buffer $\mathcal{B}$ (Lines 7-9). Once the number of $\mathcal{G}^{'}_t$ reaches a threshold value $B_{D}$, we stack the GNN model with $a^{(2)}_t$ layers, and generate the positive and negative pair for the $b$-th target user in $\mathcal{B}$. Then we train the model together with the self-supervised learning mechanism described in Section \ref{sec:method:self_supervised} (Lines 11-15). After the stacked GNN model training, we valid it on the validation dataset to get the reward $r_t$ (Line 18) and store the corresponding transition into memory $\mathcal{M}^{(1)}$ and $\mathcal{M}^{(2)}$ (Lines 20-21).

To optimize the ${\pi}_1$, we fetch batches of transitions from $\mathcal{M}^{(1)}$. For a specific transition $T^{(1)}_t=(s_t,a^{(1)}_t,s_{t+1},r_t)$, we use the Q-network to select the best action $a^{(1)}_{t+1}$ for the state $s_{t+1}$ and use the target network to estimate the target value  $Q^{target}=r+\gamma {\rm{max}}_{a_{t+1}}Q(s_{t+1},a_{t+1};\Theta^{target})$. Then we use nearest neighbor mechanism to search the nearest neighbor of state-action pair $(s_t,a^{(1)}_t)$ and add its reward value upon the value of $Q^{target}$ to obtain a new target value $\hat{Q}(s_{t},a_{t})$. Then we can optimize the ${\pi}_1$ through Eq.\ref{eq:11}. This method is also applied to optimize ${\pi}_2$ (Lines 22-24).
Eventually, we retrain the GNN with the help of the trained policies ${\pi}_1$ and ${\pi}_2$ (Line 26).
The final embedding $z_i$ of each targeted user will be produced by the last attention layer and used for the downstream classification task.

We combine the the loss function of GNN and the self-supervised loss $\mathcal{L}_{pretext}$ in Eq.\ref{eq:12}.
The loss $\mathcal{L}$ of \RoSGAS is defined as follows:
\begin{equation}
    \mathcal{L}= \sum\limits_{i=1}^n(-\log \left (y_i \cdot \sigma \left(MLP \left(z_i \right) \right) \right) + \mathcal{L}_{pretext_i}) + \lambda\parallel\Theta\parallel_2,  \label{eq14}
\end{equation}
where the first term represents the cross-entropy loss function and $\parallel\Theta\parallel_2$ is the $L2$-norm of GNN model parameters, $\lambda$ is a weighting parameter. $MLP$ reduce the embedding dimension of $z_i$ to the number of categories.

\section{experimental setup}
\label{sec:exp_setup}

\subsection{Software and Hardware}
We implement \RoSGAS with Pytorch 1.8.0, Python 3.8.10, PyTorch Geometric \cite{Fey/Lenssen/2019} with sparse matrix multiplication. All experiments are executed on a sever with an NVIDIA Tesla V100 GPU, 2.20GHz Intel Xeon Gold 5220 CPU with 64GB RAM. The operating system is Ubuntu 16.04.6.
To improve the training efficiency and avoid overfitting, we employ the mini-batch technique to train \RoSGAS and other baselines. 

\subsection{Datasets}
\label{sec:exp_setup:dataset}
We build heterogeneous graph for experiments based upon five public datasets.
The detailed description of these datasets is as follows:
\begin{itemize}
\item Cresci-15~\cite{cresci2015fame} encompasses two types of benign accounts, including a) TFP, a mixture of account set from researchers and social media experts and journalists, and b) E13, an account set consists of particularly active Italian Twitter users. Three types of social bots were collected from three different Twitter online markets, called FSF, INT, TWT. 
\item Varol-17~\cite{varol2017online} collects 14 million accounts of Twitter during three months in 2015. 3,000 accounts are sampled and selected according to some given rules. These accounts are then manually annotated into benign accounts and bot accounts. 
\item Vendor-19~\cite{yang2019arming} include a collection of fake followers deriving from several companies. To create a mixture of benign and bot accounts, we mix Vendor-19 with Verified \cite{yang2020scalable} that contains benign accounts only. 
\item Cresci-19~\cite{mazza2019rtbust} contains the accounts that are associated with Italian retweet, collected between 17-30 June 2018.
\item Botometer-Feedback~\cite{yang2019arming} stems from social bot detector Botometer. The dataset contains manually-annotated accounts based on the feedback from Botometer.
\end{itemize}

The statistics of these datasets are outlined in Table~\ref{tab:dataset-statistic}.
The number of each class of labelled nodes in each data set are basically balanced.

\begin{table}[t]
\centering
\begin{tabular}{c|cccccc}
\hline
\toprule
\specialrule{0em}{1pt}{1pt}
Dataset   & Nodes   & Edges & Benign & Bots  & Labels & Un-Labels \\ 
\specialrule{0em}{1pt}{1pt}
\hline
\specialrule{0em}{1pt}{1pt}
Cresci-15 & 2,263,472 & 10,782,235 & 1,950 & 3,339 & 5,289    & 99.77\% \\
Varol-17  & 1,978,967 & 4,916,116  & 1,244 & 639  & 1,883   & 99.90\% \\
Vendor-19 & 3,208,255 & 11,479,317 & 1,893 & 569  & 2,462    & 99.92\% \\
Cresci-19 & 669,616   & 3,341,084  & 269  & 297  & 566      & 99.92\% \\
Botometer-Feedback & 468,536  & 1,333,762 & 276  & 82 & 358 & 99.92\% \\ 
\specialrule{0em}{1pt}{1pt}
\bottomrule
\hline
\end{tabular}
\caption{Statistics of datasets.}
\label{tab:dataset-statistic}
\end{table}

\subsection{Feature Extraction}
\label{sec:exp_setup:feature_extraction}

The original datasets above merely include the metadata such as age, nickname, etc. and the posted tweet of the social account. This information, however, is insufficient to construct the heterogeneous graph, required for the effective subgraph embedding. 
These publicly released datasets originally included the social accounts' metadata (e.g., accounts' age, nickname) and the account's posted tweet data.
Since these public data are not enough to construct the heterogeneous graph we designed, we use twitter APIs to further crawl and obtain the 
the metadata and tweet data of the friends and followers of the original accounts. We form these nodes into a huge heterogeneous social graph via the multiple relationships aforementioned in Fig.~\ref{fig:meta-path}. We then use the NLP toolkit spaCy to extract name entities and treat them as a type of node in the heterogeneous graph.

In addition, we extracted the original feature vector for each type of node in the heterogeneous graph and further explored the following information as additional features: 

\begin{itemize}
    \item \textit{Account nodes:} We embed the \textit{description} field of the account metadata into a 300-dimension vector through the pre-trained language model Word2Vec. We also extracted some features such as \textit{status}, \textit{favorites}, and \textit{list} field that would be helpful for bot detection according to \cite{yang2020scalable}. We  count the number of followers and the number of friends as the key  account features, because the number of followers and friends are the most representative of a Twitter user, and using them could more efficiently and accurately describe a Twitter user. In addition, we divide the numbers by the user account lifetime (i.e., the number of years since the account was created) to reflect the changes during the whole life-cycle of an account. We also use boolean value to flag the fields including \textit{default\_profile}, \textit{verified} and \textit{profile\_use\_background\_image}, count the length of \textit{screen\_name}, \textit{name} and \textit{description} fields and the number of digits in \textit{screen\_name} and \textit{name} fields. We combine all these values as the original node features.
    
    \item \textit{Tweet nodes}: We embed the text of the original tweet by the pre-trained language model Word2Vec into a 300 dimension vector. Apart from the original node features, we further combine additional information -- the number of retweets, the number of replies, the number of favorites, the number of mentioning of the original tweet, and the number of hashtags and the number of URLs involved in the tweet.
   
    \item \textit{Hashtag and entity nodes:}  we also embed the text of hashtag and entity into a 300-dimension vector. We use zero to fill the blank holes if the number of dimension is less than 300.
\end{itemize}

The graph constructed by each data set contains millions of edges and nodes, which greatly increases the difficulty of the social bot detection task.
Noticeably, most samples in the datasets are unlabelled, i.e., more than 99\% samples are not annotated.

\subsection{Baselines and Variations}
\label{sec:exp_setup:baselines}

\subsubsection{Baselines}
\label{sec:exp_setup:baselines:baselines}

To verify the effectiveness of our proposed \RoSGAS, We compare with various semi-supervised learning baselines.
Because these baselines will run on very large-scale graphs, to ensure training and inference on limited computing resources, we use the PyG \cite{Fey/Lenssen/2019} which calculation on sparse matrix multiplication to implement these baselines.
The detail about these baselines as described as follows:

\begin{itemize}
\item \textbf{Node2Vec} \cite{grover2016node2vec} is built upon DeepWalk and introduces two extra biased random walk methods, BFS and DFS.  Compared with the random walk without any guidance, Node2Vec sets different biases to guide the procedure of the random walk between BFS and DFS, representing the structural equivalence and homophily at the same time.

\item \textbf{GCN} \cite{GCN} is a representative of the spectral graph convolution method. It uses the first-order approximation of the Chebyshev polynomial to fulfill an efficient graph convolution architecture. GCN can perform convolutional operation directly on graphs.  

\item \textbf{GAT} \cite{GAT} is a semi-supervised homogeneous graph model that utilizes attention mechanism for aggregating
neighborhood information of graph nodes. GAT uses self-attention layers to calculate the importance of edges and assign different weights to different nodes in the neighborhood. GAT also employs a multi-head attention to stabilize the learning process of self-attention. 

\item \textbf{GraphSAGE} \cite{GraphSAGE} is a representative non-spectrogram method.
For each node, it samples neighbors in different hops for the node and aggregates the features of these neighbors to learn the representation for the node. GraphSAGE improves the scalability and flexibility of GNNs.

\item \textbf{GraphSAINT} \cite{zeng2019graphsaint} is an inductive learning method based on graph sampling. It samples subgraphs and performs GCN on them to overcome the neighbor explosion problem while ensuring unbiasedness and minimal variance.

\item \textbf{SGC} \cite{wu2019simplifying} is a simplified graph convolutional neural network. It reduces the excessive complexity of GCNs by repeatedly removing the non-linearity between GCN layers and collapsing the resulting function into a single linear transformation. This can ensure competitive performance when compared with GCN and significantly reduce the size of parameters.

\item \textbf{ARMA} \cite{bianchi2021graph} is a non-linear and trainable graph filter that generalizes the convolutional layers based on polynomial filters. ARMA can provide GNNs with enhanced modeling capability.

\item \textbf{Policy-GNN} \cite{lai2020policy} is a meta-policy framework that adaptively learns an aggregation policy to sample diverse iterations of aggregations for different nodes. It also leverages a buffer mechanism for batch training and a parameter sharing mechanism for diminishing the training cost.
\end{itemize}

\subsubsection{Variants} 
\label{sec:exp_setup:baselines:variants}

We generate several variants of the full \RoSGAS model, to more comprehensively understand how each module works in the overall learning framework and better evaluate how each module individually contribute to the performance improvement.  \RoSGAS mainly comprises three modules: Reinforced Searching Mechanism, Nearest Neighbor Mechanism, and self-supervised learning.  We selectively enable or disable some parts of them to carry out the ablation study. 

The details of these variations are described as follows:
\begin{itemize}
\item \textbf{\RoSGAS-\textit{K}:} This variant only enables the reinforced searching, without the aid of any other modules, to find out the width ($k$) of the subgraph for every target user $v_i$. Due to the huge scale of the constructed social graph, the search range will be limited to $[1,2]$ to prevent the explosion of neighbors. Nevertheless, such search range can be flexibly customized to adapt to any other scenarios and datasets. In the context of this model variant, the number of layers is fixed to be $l=3$.  

\item \textbf{\RoSGAS-\textit{L}:} This variant only switches on the reinforced searching mechanism for pinpointing the number of the layers ($l$) to stack the GNN model for every target user $v_i$. The search range will be limited to $[1,3]$ to save computing resources. The width $k$ of the subgraph is fixed $k=2$. 

\item \textbf{\RoSGAS-\textit{KL}:} This variant enables both the subgraph width search and the layer search in the reinforced searching mechanism for each target user $v_i$. In this model variant, the width is set to be $k\in[1,2]$ while the number of layers of the GNN model is set to be $l\in[1,3]$. 

\item \textbf{\RoSGAS-\textit{KL}-\textit{NN}}: This variant utilizes the reinforced search mechanism, together with the nearest neighbor mechanism. In the optimization process of the RL agent, the nearest neighbor module can stabilize the learning in the early stages of RL as soon as possible, and accelerate the model convergence.

\item \textbf{\RoSGAS} contains all modules in the learning framework. The self-supervised learning mechanism is additionally supplemented upon \RoSGAS-\textit{KL}-\textit{NN}.

\end{itemize}

\subsection{Model Training}
We use the following setting: embedding size (64), batch size (64), the base layer of \RoSGAS (GCN), learning rate (0.05), optimizer (Adam), L2 regularization weight ($\lambda_2=0.01$), and the training epochs (30). As aforementioned in Section \ref{sec:exp_setup:baselines:variants}, we set the action (range) of searching GNN layers from 1 to 3 and the action (range) of subgraph width searching from 1 to 2 to prevent neighbors from exploding 
for the DQN \cite{dqn}. The agent training episodes is set to be 20 and we construct 5-layer of MLP with 64, 128, 256, 128, 64 hidden units. 
We use the accuracy obtained from the validation set to select the best RL agent and compare the performance with the other models in the test set.
As for the nearest neighbor mechanism, we set the ($\rm{L}=7$), the initial $\alpha_0=0.5$.

\subsection{Evaluation Metrics}

As the number of labelled benign accounts and malicious accounts in the several data sets used is well-balanced, we utilize Accuracy to indicate the overall performance of classifiers:

\begin{equation}
Acurracy=\frac{TP+TN}{TP+FP+FN+TN},  \label{eq:13}
\end{equation}
where $TP$ is True Positive, $TN$ is True Negative, $FP$ is False Positive, $FN$ is False Negative.
\section{Experiment Results}
\label{sec:exp_res}

In this section, we conduct several experiments to evaluate \RoSGAS. We mainly answers the following questions: 

\begin{itemize}
\item \textbf{Q1:} How different algorithms perform in different scenarios, i.e., algorithm effectiveness (Section~\ref{sec:exp_res:accuracy}).

\item \textbf{Q2:} How each individual module of \RoSGAS contributes to the overall effectiveness  (Section~\ref{sec:exp_res:ablation}).

\item \textbf{Q3:} How the RL search mechanism work in terms of effectiveness and explainability (Section~\ref{sec:exp_res:rl}).

\item \textbf{Q4:} How fast different algorithm can achieve, i.e., efficiency (Section~\ref{sec:exp_res:efficiency}) and how the RL algorithms can converge effectively (Section ~\ref{sec:exp_res:convergence}).

\item \textbf{Q5:} How different algorithms perform when dealing with previously unseen data, i.e., generalization (Section~\ref{sec:exp_res:generalization}).

\item \textbf{Q6:} How to explore the detection result and visualize the high-dimensional data  (Section~\ref{sec:exp_res:casestudy}).
\end{itemize}

\subsection{Overall Effectiveness}
\label{sec:exp_res:accuracy}

In this section, we conduct experiments to evaluate the accuracy of the social bot detection task on the five public social bot detection datasets.
We report the best test results of baselines,  \RoSGAS, and the variants. We performed 10-fold cross-validation on each dataset.
As shown in Table~\ref{tab:comparision-table}, \RoSGAS outperforms other baselines and different variants under all circumstances. This indicates the feasibility and applicability of \RoSGAS in wider ranges of social bot detection scenarios.
Compared with the best results among all the state-of-the-arts, our method can achieve 3.38\%, 9.55\%, 5.35\%, 4.86\%, 1.73\% accuracy improvement, on the datasets of Cresci-15, Varol-17, Vendor-19, Cresci-19 and Botometer-Feedback, respectively.

In the baselines, Node2Vec is always among the worst performers in the majority of datasets. This is because Node2Vec controls the random walk process by setting a probability p to switch between the BFS and the DFS strategy. Node2Vec sometimes fails to obtain the similarity of adjacent nodes in large-scale social graph with extremely complex structures, and does not make good use of node features.  GraphSAGE samples the information of neighbors for the aggregation for each node. This design can not only reduce information redundancy but also increase the generalization ability of the model through randomness. However, the proposed random sampling is not suited for super large-scale graph, and the inability to adapt to the change of the receptive field drastically limit its performance. GCN multiplies the normalized adjacency matrix with the feature matrix and then multiplies it with the trainable parameter matrix to achieve the convolution operation on the whole graph data. However, obtaining the global representation vector for a node through full-graph convolution operations would massively reduce the generalization performance of the model. Meanwhile, the increases of receptive field will lead to a soaring number of neighbors, and thus weaken the ability of feature representation.
GAT shares weight matrix for node-wise feature transformation and then calculates the importance of a node to another neighbor node, before  aggregating the output feature vector of the central node through the weighted product and summation. GAT also experiences the explosion of receptive field and the consequent increase of the neighbor number. 
Unlike GraphSAGE, GraphSAINT sets up a sampler to sample subgraphs from the original graph. It uses GCN for convolution on the subgraph to resolve neighbor explosion and sampling probability is set to ensure unbiasedness and minimum variance. However, extracting subgraphs in GraphSAINT is random and thus limit the precision of subgraph embedding. 
SGC simplifies the conventional GCNs by repeatedly removing the non-linearities between GCN layers and collapsing the resulting function into a single linear transformation. Namely, the nonlinear activation function between each layer is removed to obtain a linear model. Compared with GCN, SGC can achieve similar performance, with a slightly-reduced accuracy among all datasets.
In addition, \RoSGAS also outperforms Policy-GNN and ARMA since these counterparts merely make exclusive improvement on convolution layers. By comparison,  our approach takes advantage of subgraph search and GNN architecture search, whilst leveraging self-supervised learning to overcome the limitation of limited labelled samples. These functionalities can fulfill better performance even only based upon the basic GCN layer.

\begin{table*}[t]
\centering
\scalebox{0.95}{
\begin{tabular}{c|ccccccc}
\hline
\toprule
Method    & Cresci-15 & Varol-17  & Vendor-19 & Cresci-19 & Botometer-Feedback \\ 
\specialrule{0em}{1pt}{1pt}
\hline
\specialrule{0em}{1pt}{1pt}
Node2Vec \cite{grover2016node2vec}   & 73.02$\pm$0.91 & 61.43$\pm$0.8 & 76.13$\pm$2.61 & 76.65$\pm$4.97 & 75.68$\pm$0.81         \\
GraphSAGE \cite{GraphSAGE}  & 91.94$\pm$1.12 & 65.71$\pm$1.62 & 81.65$\pm$1.19 & 70.43$\pm$0.31 &  76.16$\pm$0.43         \\
\specialrule{0em}{1pt}{1pt}
\hline
\specialrule{0em}{1pt}{1pt}
GCN \cite{GCN} & 94.22$\pm$0.57 & 65.15$\pm$1.85 & 85.84$\pm$0.81 & 72.75$\pm$2.65 & 76.02$\pm$0.34          \\
GAT \cite{GAT}& 94.05$\pm$0.31 & 64.63$\pm$1.45 & 78.04$\pm$0.78 & 71.74$\pm$1.63 & 75.21$\pm$1.46          \\
SGC \cite{wu2019simplifying} & 89.32$\pm$0.65 & 63.32$\pm$7.24 & 78.71$\pm$0.75 & 64.34$\pm$1.67 &  74.51$\pm$0.35 \\
GraphSAINT \cite{zeng2019graphsaint} & 80.93$\pm$0.92 & 64.89$\pm$1.26 & 77.60$\pm$2.51 & 78.55$\pm$1.14 &  75.51$\pm$0.59         \\
Policy-GNN \cite{lai2020policy} & 93.44$\pm$0.13 & 66.20$\pm$4.21 & 82.01$\pm$0.23 & 80.19 $\pm$7.35 & 76.61$\pm$1.17\\
ARMA \cite{bianchi2021graph}& 94.71$\pm$0.43 & 65.43$\pm$3.24 & 83.44$\pm$1.94 & 78.46$\pm$1.52 & 75.77$\pm$0.71\\
\specialrule{0em}{1pt}{1pt}
\hline
\specialrule{0em}{1pt}{1pt}
\RoSGAS-$K$    & 93.13$\pm$0.09 & 66.06$\pm$0.19 & 76.88$\pm$ 0.78& 80.38$\pm$0.29 & 77.09$\pm$0.42 \\
\RoSGAS-$L$    & 97.82$\pm$0.10 & 67.42$\pm$0.37 & 77.40$\pm$0.45 & 82.80$\pm$0.16 & 77.25 $\pm$ 0.90  \\
\RoSGAS-$KL$ & 97.82$\pm$0.14 & 68.01$\pm$2.72 & 87.84$\pm$1.22 & 82.95$\pm$1.13 & 77.09$\pm$ 0.14           \\ 
\RoSGAS-$KL$-$NN$ & 97.84$\pm$0.05 & 75.11$\pm$0.07 & 90.52$\pm$0.03 & 84.88$\pm$0.01 & 77.93$\pm$0.14           \\ 
\RoSGAS      & \textbf{98.09}$\pm$\textbf{0.36} & \textbf{75.75}$\pm$\textbf{0.31} & \textbf{91.19}$\pm$\textbf{0.35} & \textbf{85.05}$\pm$\textbf{0.21} & \textbf{78.34}$\pm$\textbf{0.25}         \\ 
\specialrule{0em}{1pt}{1pt}
\hline
\specialrule{0em}{1pt}{1pt}
Gain      & \textbf{3.38} $\uparrow$             & \textbf{9.55} $\uparrow$             & \textbf{5.35} $\uparrow$             & \textbf{4.86} $\uparrow$           &  \textbf{1.73} $\uparrow$ \\
\bottomrule
\hline
\end{tabular}
}
\caption{Comparison of the average accuracy of different methods for social bot detection (unit: \%).}
\label{tab:comparision-table}
\end{table*}

\subsection{Ablation Study}
\label{sec:exp_res:ablation}

The second half of Table~\ref{tab:comparision-table} also compares the performance of multiple variants to demonstrate how to break down the performance gain into different optimization modules.

While \RoSGAS-$K$ only uses the RL agent for subgraph width search, the achieved effectiveness is competitive against the state-of-the-arts in some datasets, such as Varol-17, Cresci-19, and Botometer-Feedback. \RoSGAS-$L$ that only enables the architecture search can even achieve better accuracy than other baselines in most datasets, except the vendor-19 dataset. These observations are in line with the enhancement provided by the RL capability. Intrinsically, putting the searching mechanisms together will bring synergetic benefits to the effectiveness. Particularly, for datasets Vendor-19, the combination of width search and layer search can significantly make \RoSGAS outstanding from the other counterparts, reaching 87.84\% accuracy on average. We can also observe that the contribution  of layer search to the performance gain appears to be more significant than the width search. This is because the embedding effectiveness is less insensitive to the neighbor selection, and the number of GNN layers can more effectively affect the overall effectiveness. The increased layer can make the nodes closer to the target node more frequently aggregated to form the embedding of the target node. Likewise, nodes far from the target node will be less involved in the embedding. Hence, an enhanced embedding effectiveness can be gained from enabling GNN layer search. These findings indicate the necessity of jointly searching the width of subgraphs and the number of GNN layers to unlock the full potential of performance optimization.
 
As shown in the result of \RoSGAS-KL-NN,  integrating the nearest neighbor mechanism with the backbone searching mechanism can further enhance the performance gain stemming from the learning process in the RL agent. Most noticeably, the accuracy can be substantially augmented from 68\% to 75\% by the nearest neighbor design when tackling Varol-17 dataset. Even if in some cases, the improvement is not as significant as that in Varol-17, the similarity driven nearest neighbor scheme can boost the action exploration in the RL agent and thus help with accuracy promotion. 
Furthermore, we demonstrate an incremental performance gain from adopting the proposed  self-supervised learning mechanism. Across all different datasets, up to 0.7\% improvement can be further achieved despite marginal increment observed in some datasets. Further investigation would be required to better leverage large quantities of unlabeled data and enhance the generalization ability of the self-supervised classifiers in different scenarios. This is beyond the core scope of this paper and will be left for future study.

\begin{figure}[t]
\centering
\includegraphics[width=0.83\textwidth]{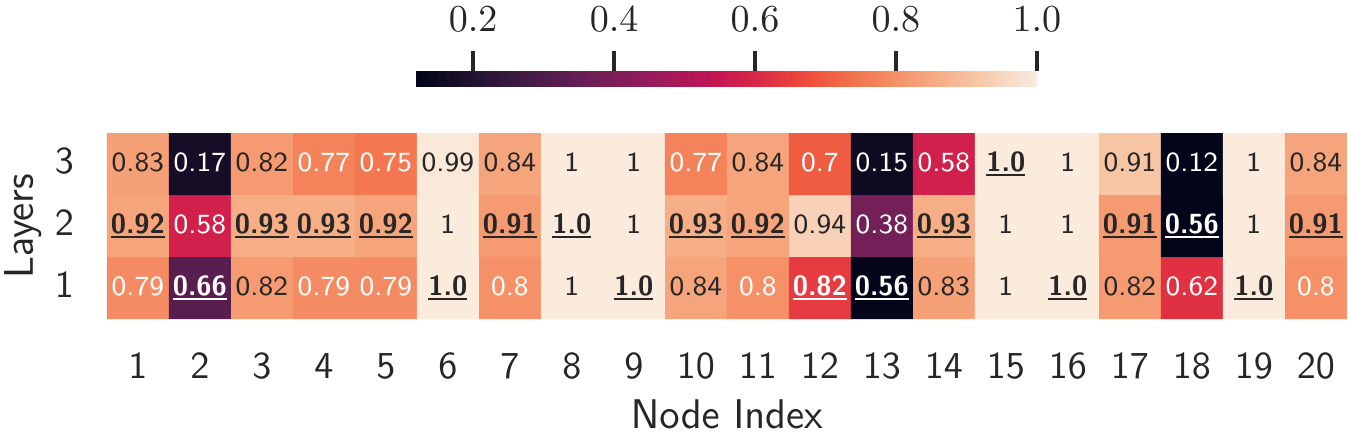}
\caption{Per-node prediction effectiveness when conducting architecture search. The underline represents the selected layer number  by the RL-agent in the \RoSGAS for each node.}
\label{fig:heatmap}
\end{figure}

\begin{figure}[t]
\centering
\subfigure[The upper bound of width of subgraph]{
\label{tab:parameter-sensitivity-sub-1}
\includegraphics[width=0.45\textwidth]{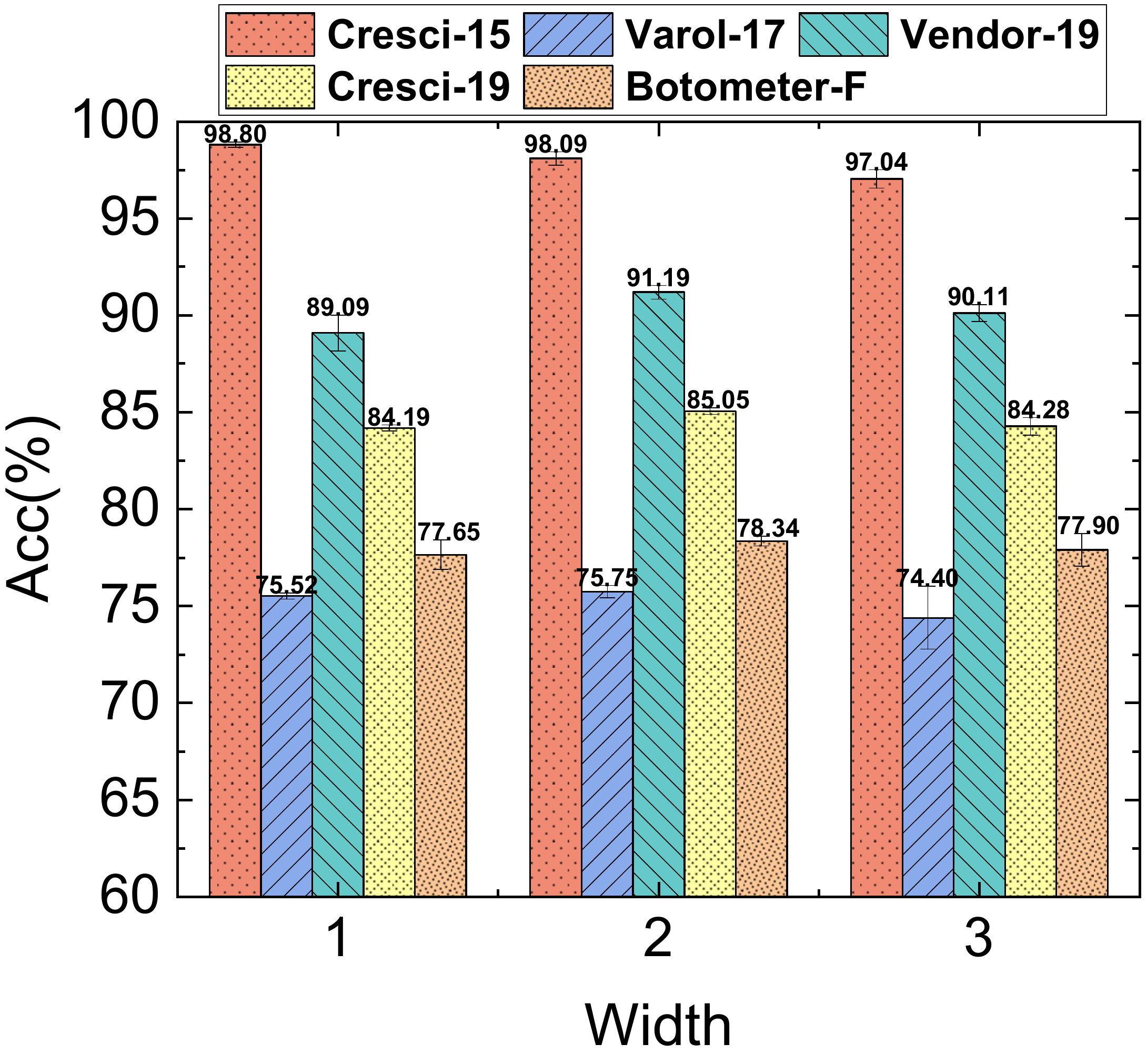}}
\subfigure[The upper bound of the number of GNN layers.]{
\label{tab:parameter-sensitivity-sub-2}
\includegraphics[width=0.45\textwidth]{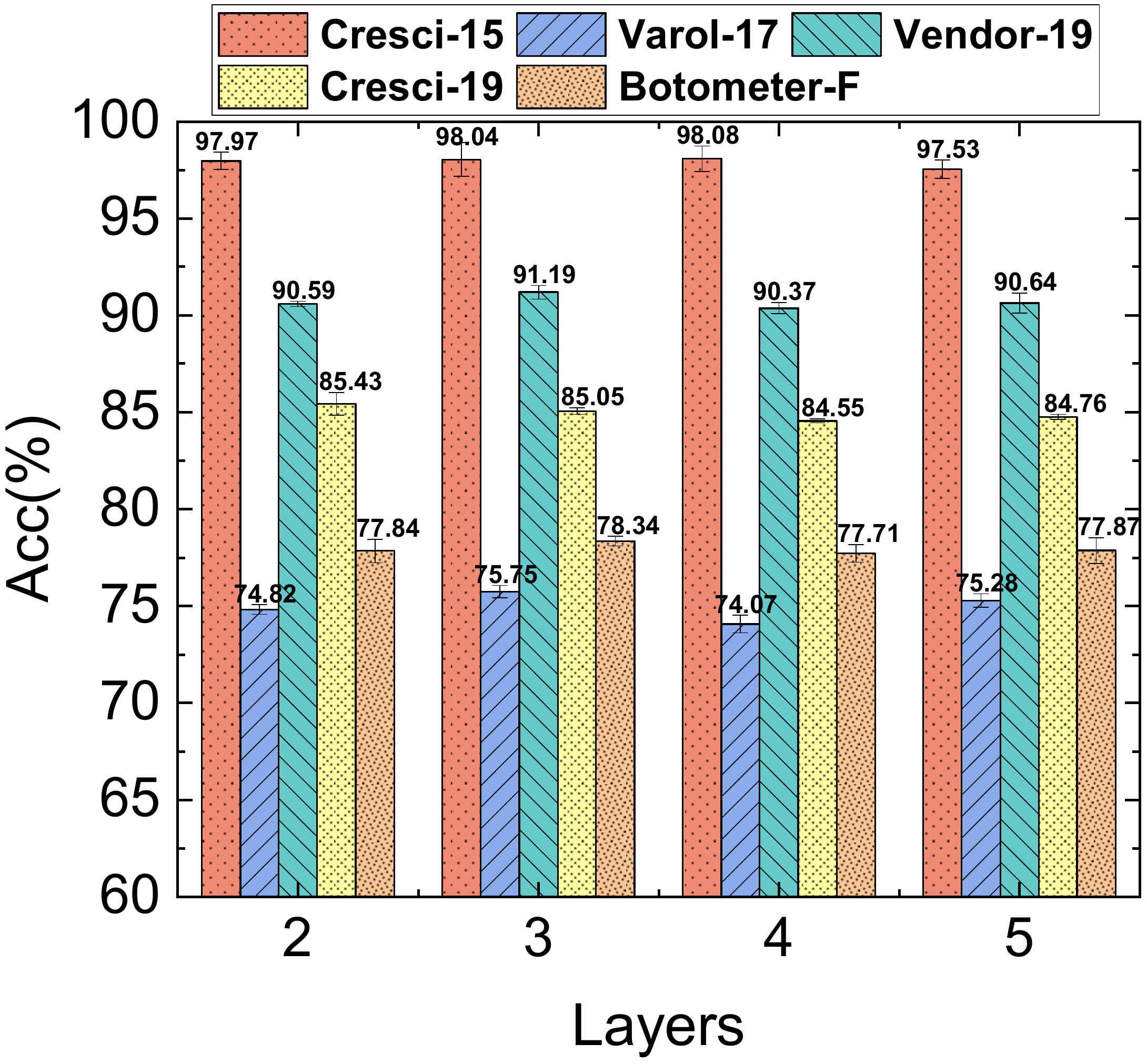}}
\caption{Impact of search upper bound of subgraph width and layer number on the effectiveness}
\label{fig:parameter-sensitivity}
\end{figure}

\subsection{Effectiveness of the RL Mechanism}
\label{sec:exp_res:rl}

\subsubsection{Parameter Searching Result by the RL Agent}

To conduct an in-depth investigation in how effectiveness of the RL search mechanism, we randomly select 20 labeled nodes from the Cresci-15 dataset as the target users and examine the accuracy when adopting some given circumstances of GNN layers. For example, we create 3 types of GNN models, by manually stacking 1 layer of GCN, 2 layers of GCN, and 3 layers of GCN, respectively. 
We feed the same graph data used in the previous testing into the three models and train 100 times. The trained models are used for validating if the RL can obtain the most proper number of layers for the selected nodes. For each node in the 20 target users, we count the ratio of being correctly classified out of the 100 runs.
The main purpose of this investigation is to examine if the RL-based mechanism can pick up the layer with the highest classification ratio, to automatically enable the best model performance.  

Fig.~\ref{fig:heatmap} shows the ratio obtained for each target node when using different layers. Observably, different GCN layers have a varying impact on the correct prediction of a certain node. 
For example, the prediction effectiveness of other nodes (e.g.,  node index 2, 4, 5, 11, 12, 13 and 18 ) will be drastically affected by the number of GNN layers. We reason this phenomenon is because the range of GNN's receptive field will gradually increase when the layer number ramps up; meanwhile, higher order neighbor information will be involved and aggregated, thereby having a direct impact on the detection accuracy. By contrast, some nodes (e.g., node index 6, 8, 9, 15, 16 and 19) can be better predicted no matter what information is aggregated from neighbors with different orders.  
It is therefore necessary to elaborately select the number of GNN layers for these nodes to increase the probability of correct prediction of these nodes.

We use underline, e.g., \underline{\textbf{0.92}}, to mark the final decision made by the RL agent when searching the model layer. The proposed RL agent can select the optimal layer that can deliver the highest prediction ratio.
There is only 10\% (2 of 20 classification tasks) mismatch between the best layer option and the choice made by the RL agent. This indicates the proposed approach can effectively reduce the manual tuning whilst reaching the best effectiveness.

\subsubsection{Impact of different search ranges on the effectiveness}

We dive into the impact of parameter selection on the overall effectiveness and demonstrate the sensitivity to such parameter changes. 
To do so, we first fix the maximum searching bound of the number of search layers of the graph neural network to be 3, whilst gradually increasing the searching bound for the width of the subgraph from 1 to 3.    As shown in Fig.~\ref{tab:parameter-sensitivity-sub-1}, for all datasets without exception, all the model instances experience a rise of accuracy when the searching range of width grows to 2 but a slight drop when the range is further extended to 3. 
This is because the increase of width range will lead to a hugely growing number of neighbors involved in the extracted subgraph.  Particularly  the scale of the constructed graphs in the  large-scale datasets are normally large and will lead to the explosion of  neighbors, which in turn give rise to  the reduced quality of graph embedding and lower accuracy. This observation indicates the search range for extracting subgraphs needs to be carefully modified and adaptive to the scale of a given scenario. 

Given the best result can be stably obtained by adopting $[1,2]$ as the width range, we then fix this setting and varying the the search range of the number of model layers. We gradually ramp up the upper bound of the rang from 2 to 5.  As shown in Fig.~\ref{tab:parameter-sensitivity-sub-2}, there is no significant disparities 
in the effectiveness among different options. The effectiveness is insensitive to the change of model layers despite some noticeable variations. 
For example, in the Cresci-15 dataset, the model accuracy will reach the peak when choosing 4 as the upper bound of the searching range while for Vendor-19 the accuracy peak will come when searching up to 3 layers. 
Nevertheless, the discrepancy in accuracy is negligible and the proposed RL-based searching mechanism can more flexibly and adaptively make the best decision in ensuring the best model performance without incurring excessive cost in exploring additional GNN layers.

\begin{table}[t]
\centering
\begin{tabular}{@{}c|ccccc@{}}
\hline
\toprule
Method       & Cresci-15  & Varol-17 & Vendor-19 & Cresci-19 & Botometer-Feedback\\ 
\midrule
GCN          & 572.53     & 115.04   & 323.36    & 18.78   &   10.21 \\
GAT          & 576.72     & 118.51   & 331.66    & 21.75   &   11.64\\
SGC          & 296.38     & 198.62   & 331.43    & 108.10  &   39.97 \\
GraphSAINT   & 604.13     & 167.11   &  460.13   & 55.98   &   12.56\\
Policy-GNN   & 2076.15    & 578.21   &  1734.41  & 120.34  &   109.21 \\
ARMA         & 59.79      & 51.37    &  84.89    & 18.29   &   12.61 \\
\specialrule{0em}{1pt}{1pt}
\hline
\specialrule{0em}{1pt}{1pt}
\RoSGAS-$KL$    & 587.53 & 174.12 & 354.11  & 55.87  & 26.54    \\
\RoSGAS-$KL$-$NN$  & 612.32 & 212.13 & 382.11 & 356.42 & 29.22 \\
\RoSGAS         & 809.88 & 245.62 & 469.20 & 429.6  & 32.27\\
\bottomrule
\hline
\end{tabular}
\caption{The average time consumption (unit: second) of running each method 10 times on the datasets Cresci-15, Varol-17 (Varol), Vendor-19 (Vendor), Cresci-19, Botometer-Feedback (Botometer-F), respectively.}
\label{tab:time-consumption-table}
\end{table}

\subsection{Efficiency}
\label{sec:exp_res:efficiency}

We primarily evaluate the efficiency by measuring the training time. As the reward signal needs to be obtained from the validation dataset to train the RL agent before constructing GNN stack, we break down the time consumption into two parts -- RL and GNN model training. It is worth noting that sparse matrix multiplication used in PyTorch Geometric can enable GNNs to be applied in very large graphs and accelerate the model training. 

Table~\ref{tab:time-consumption-table} presents the average time consumption of running each model 10 times.
Overall, \RoSGAS can strike a balance between the training time and the effectiveness. Although ARMA and SGC take less time to train their model due to the simplified GCN model through removing non-linearity and collapsing weight matrices between consecutive layers, the achieved accuracy is far lower than \RoSGAS across all datasets, particularly on Cresci-19 where the labels are scarcer. Policy-GNN 
takes the longest time for model training simply because the GNN stacking and convolution operation will be carried out for all nodes in the entire graph.

Most notably, the variant \RoSGAS-$KL$ which does not include the nearest neighbor mechanism and the self-supervised learning mechanism can achieve competitive training efficiency compared with GCN, GAT, and GraphSAINT, only with a marginal time increase. The slight difference is negligible when considering \RoSGAS-$KL$ needs to train both the RL agent and the GNN model separately. Compared with \RoSGAS-$KL$, \RoSGAS-$KL$-$NN$ intrinsically needs extra time to search the set of state-action pairs that have been explored and ascertain the nearest neighbor to the current state-action pair, before modifying the expectation of the reward for optimized network parameters. \RoSGAS additionally involves the self-supervised learning based on the target user batch to extract the homologous subgraphs for additional forward propagation. 

In addition, the time consumption for Cresci-15 is far larger than that for Cresci-19. There is a linearly-increased time consumption of GCN and GAT when the graph scale soars. 
In fact, the GNN training is more efficient since we only need to perform convolution operations on the sampled subgraph; by contrast, each iteration of GCN and GAT will have to perform a convolution operation on all nodes of an entire graph. 
\RoSGAS is solely relevant to the extracted number of subgraphs for detecting the target users, and thus independent from the scale of the entire graph. This clearly showcases the inherent scalability and robustness of our sample and subgraph based mechanism adopted in \RoSGAS.  
\begin{figure*}[t]
\centering
\subfigure[Cresci-15.]{
\label{tab:rl-process-sub-1}
\includegraphics[width=0.48\textwidth]{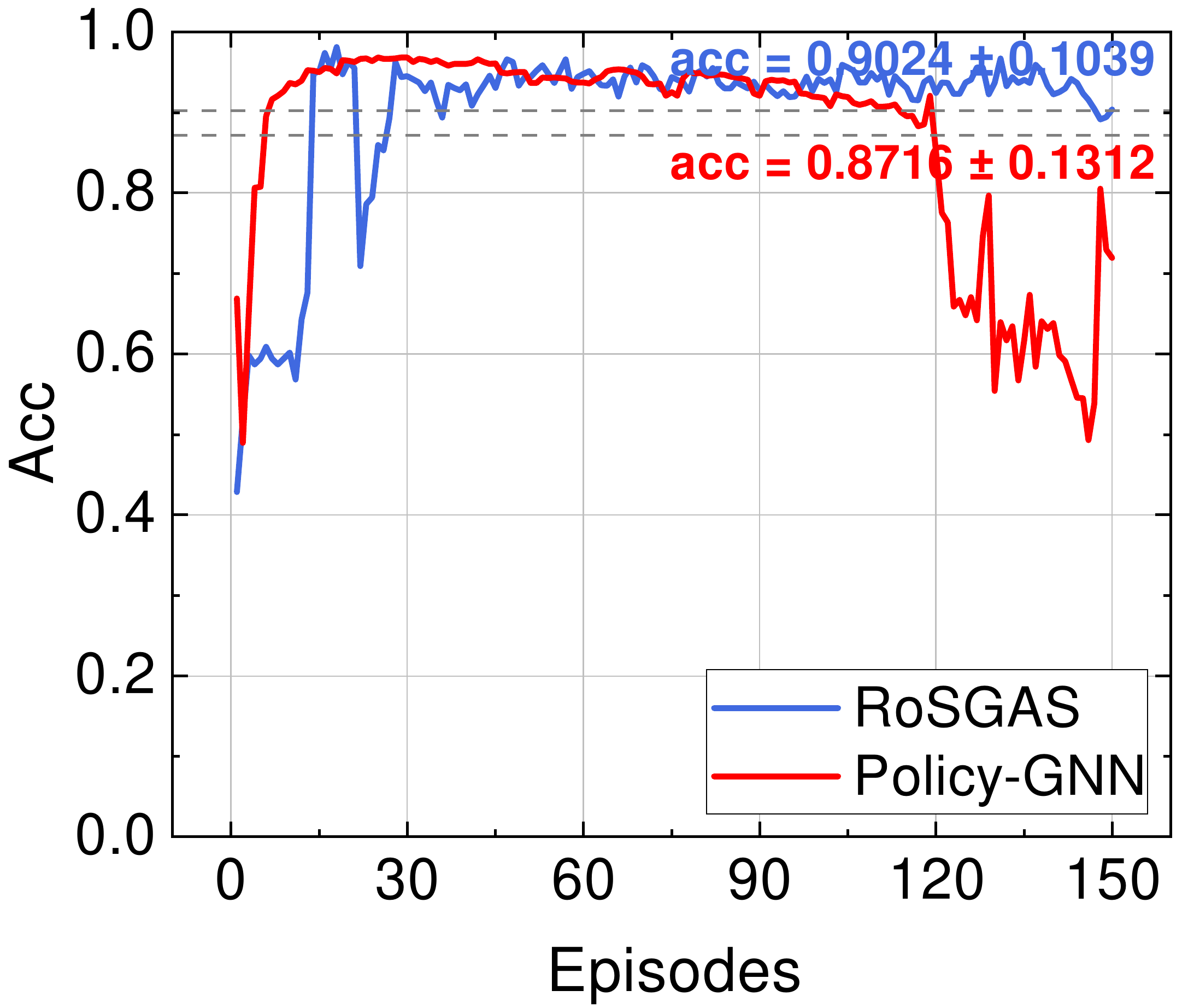}}
\subfigure[Varol-17.]{
\label{tab:rl-process-sub-2}
\includegraphics[width=0.48\textwidth]{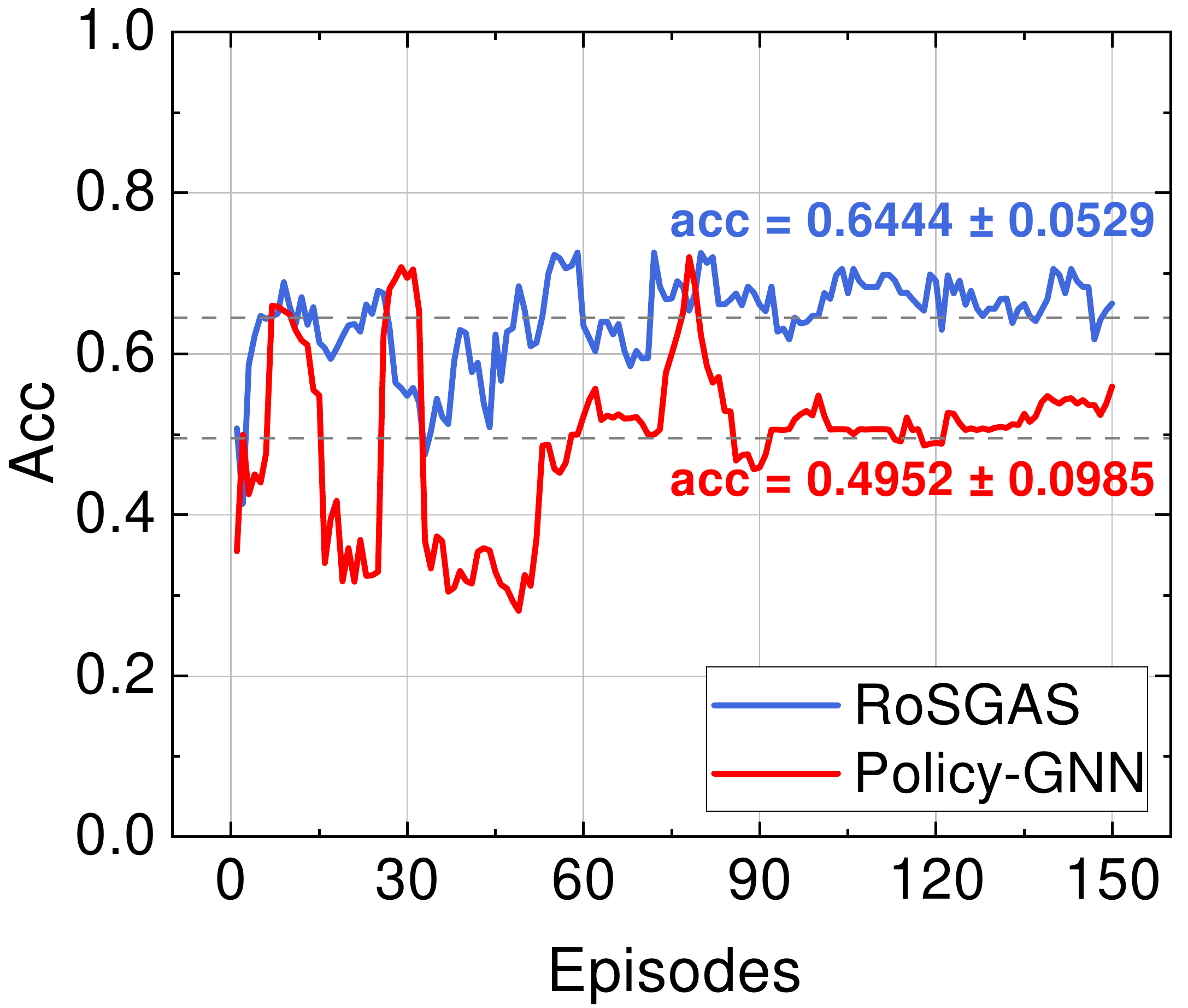}}
\subfigure[Vendor-19.]{
\label{tab:rl-process-sub-3}
\includegraphics[width=0.48\textwidth]{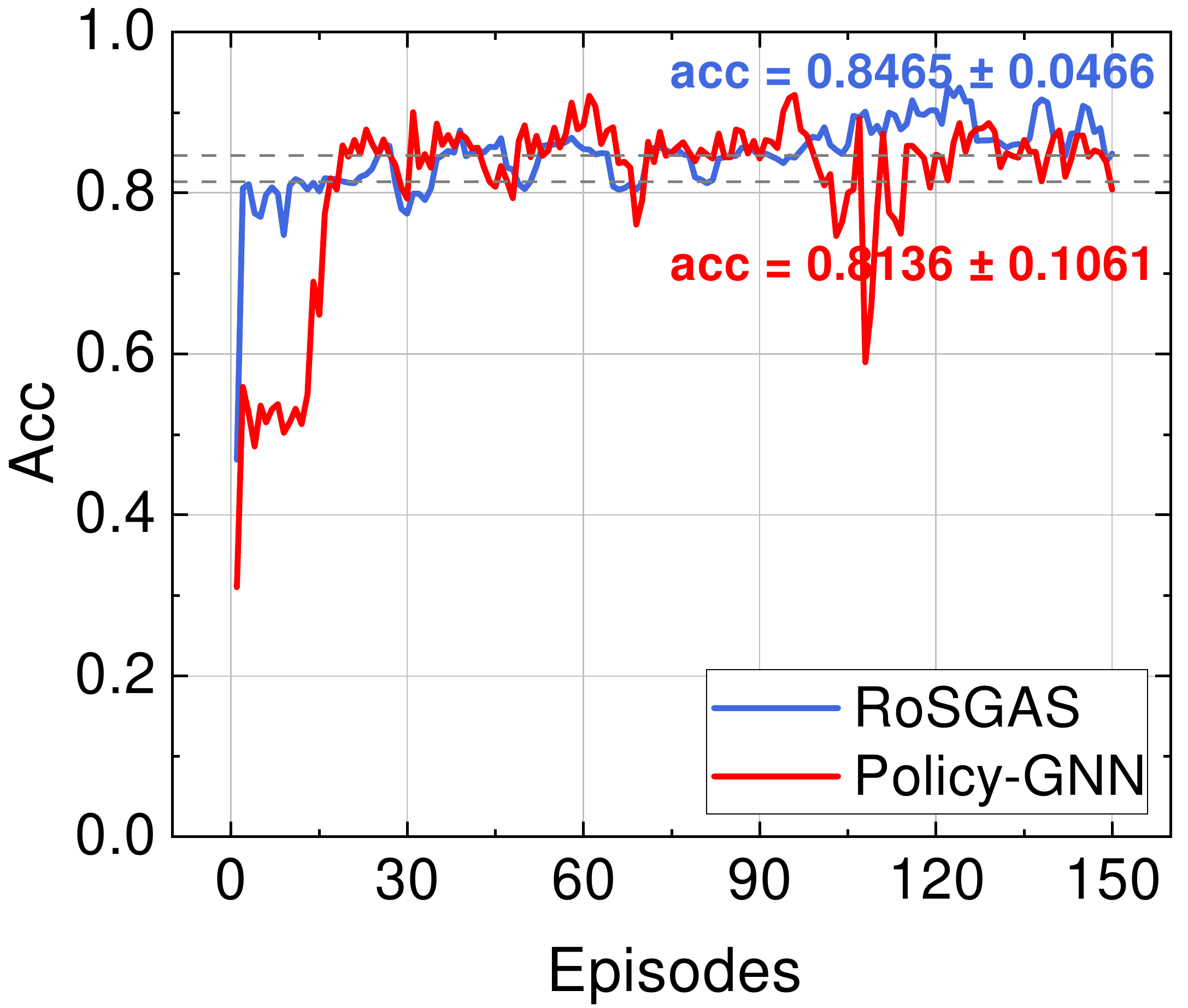}}
\subfigure[Botometer-Feedback.]{
\label{tab:rl-process-sub-4}
\includegraphics[width=0.48\textwidth]{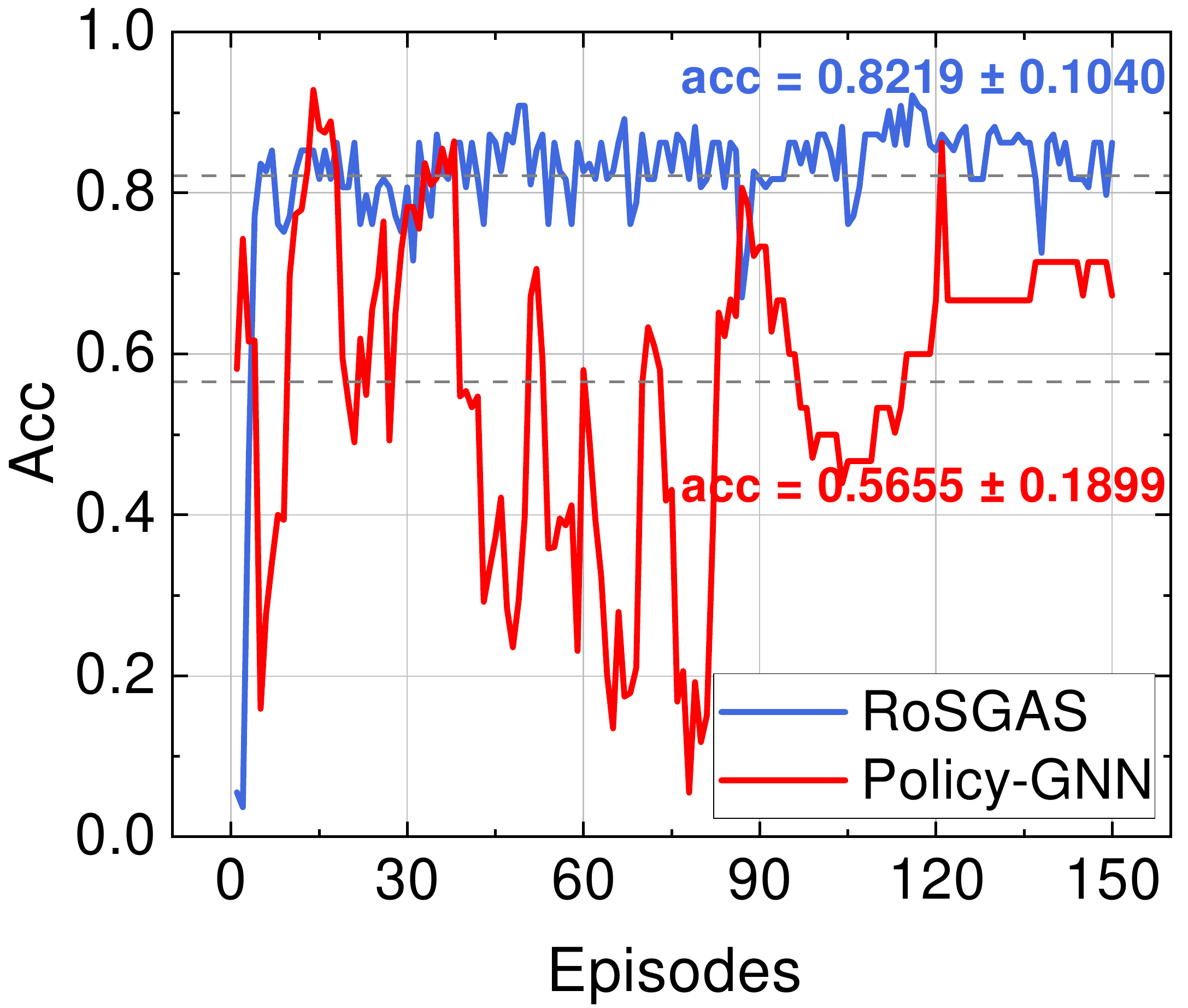}}
\caption{The RL-agent training process of \RoSGAS.}
\label{fig:RL-agent-training-process}
\end{figure*}
\subsection{Stability}
\label{sec:exp_res:convergence}

We also compare the stability of the RL-based architecture search between \RoSGAS and Policy-GNN. 
Fig.~\ref{fig:RL-agent-training-process} demonstrates the detailed training process with the RL agent on Cresci-15, Varol-17, Vendor-19 and Botometer-Feedback, respectively. The dotted line represents the accuracy obtained in the validation set during the 150 episodes. 
Obviously, \RoSGAS can promptly achieve high accuracy under all the datasets and the mean accuracy can be achieved only after 15 RL agent training episodes.
Meanwhile, once reaching this point, a stable state of Nash equilibrium can be steadily maintained without huge turbulence.

By contrast, the accuracy of Policy-GNN is noticeably lower than \RoSGAS and lacks stability, i.e., very obvious fluctuations manifest. The disparity mainly stems from the design of the state transition and the nearest neighbor mechanism. 
\RoSGAS uses the embedding of the initial subgraph as the state input to the RL agent, and jumps between the initial subgraphs as the state transition, while Policy-GNN uses the node embedding as the state input and jumps between the nodes as the state transition.  Undoubtedly, the embedding of the initial subgraph as a state can better reflect the local structure of a targeted node, resulting in a stronger representation ability, and hence the enhanced stability. At the same time, during the RL agent training episodes, for a specific state-action pair of one transition, the nearest neighbor mechanism explores the existing pairs and exploits the reward pertaining to this nearest neighbor from the environment. The reward is used as a part of the label to optimize the Q-network, which greatly eliminates the difference between the target network prediction and the actual reward in the initial stage. As a result, the RL agents can achieve higher accuracy with only a few training episodes.

Interestingly, 
\RoSGAS has different volatility across different datasets and the magnitude of volatility is positively correlated to the average correct rate. This indicates the node feature distribution and graph structure pertaining to each individual dataset have a non-trivial impact on the model training. The in-depth study will be left for future work. 
\subsection{Generalization}
\label{sec:exp_res:generalization}

In the field of social bot detection, the continuous evolution of social bots’ camouflage technology has brought generalization challenges to the model design.
To identify the emerging attack methods of social bots and mine diverse user information, a robust detection model should have a high level of generalization.
In this subsection, we examine the generalization of \RoSGAS and compare with other baselines.

\subsubsection{Out-of-sample Validation Accuracy}

Apart from the in-sample validation in previous subsections, the most effective way  to demonstrate the strong generalization is to perform out-of-sample validation,  i.e., retaining some of the sample data for model training, and then using the model to make predictions for unseen data and examine the accuracy. 
To do so, we select one dataset as the training dataset and then use other datasets as the test dataset. We divide the training set into 10 equal parts, take one part of the training set for training each time, and train each baseline for 30 epochs. We use the trained models to predict the labelled accounts in the test datasets to calculate the accuracy. 

\begin{figure}[t]
\centering
\includegraphics[width=0.97\textwidth]{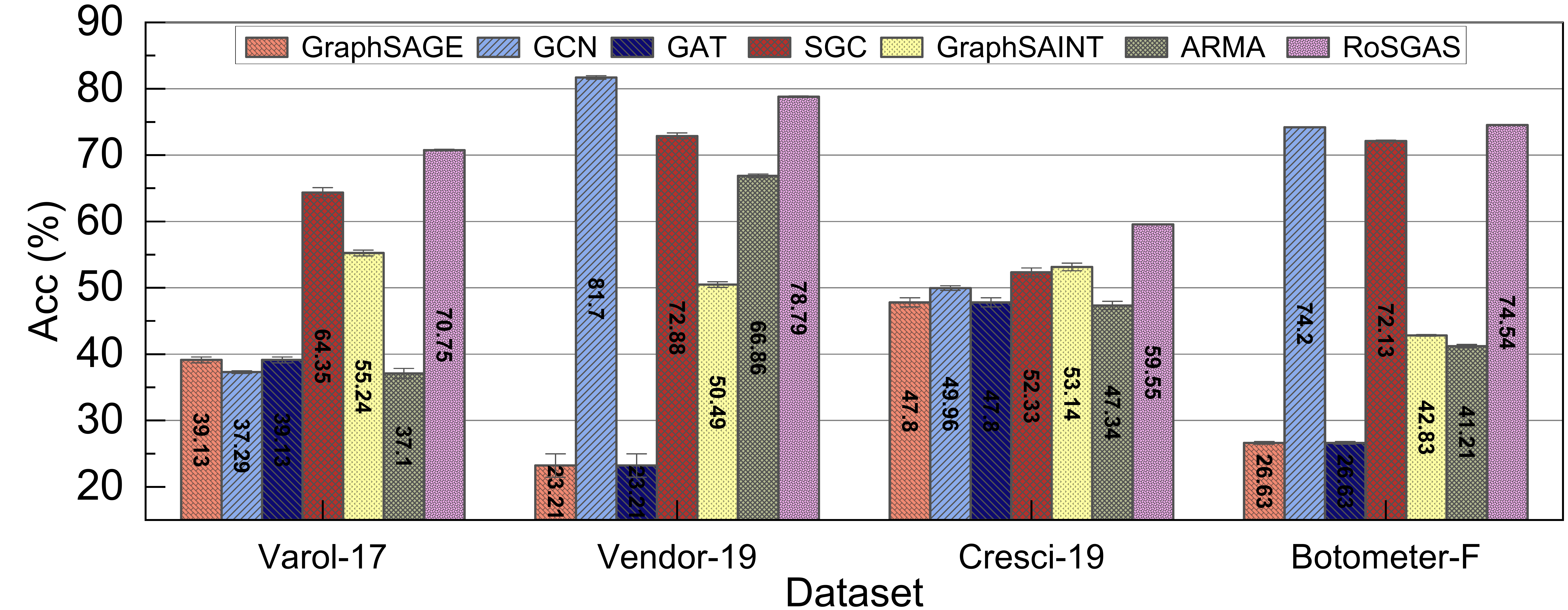}
\caption{Train on the Cresci-15 and testing on different dataset}
\label{fig:general-cresic-15}
\end{figure}

\begin{figure}[t]
\centering
\includegraphics[width=0.97\textwidth]{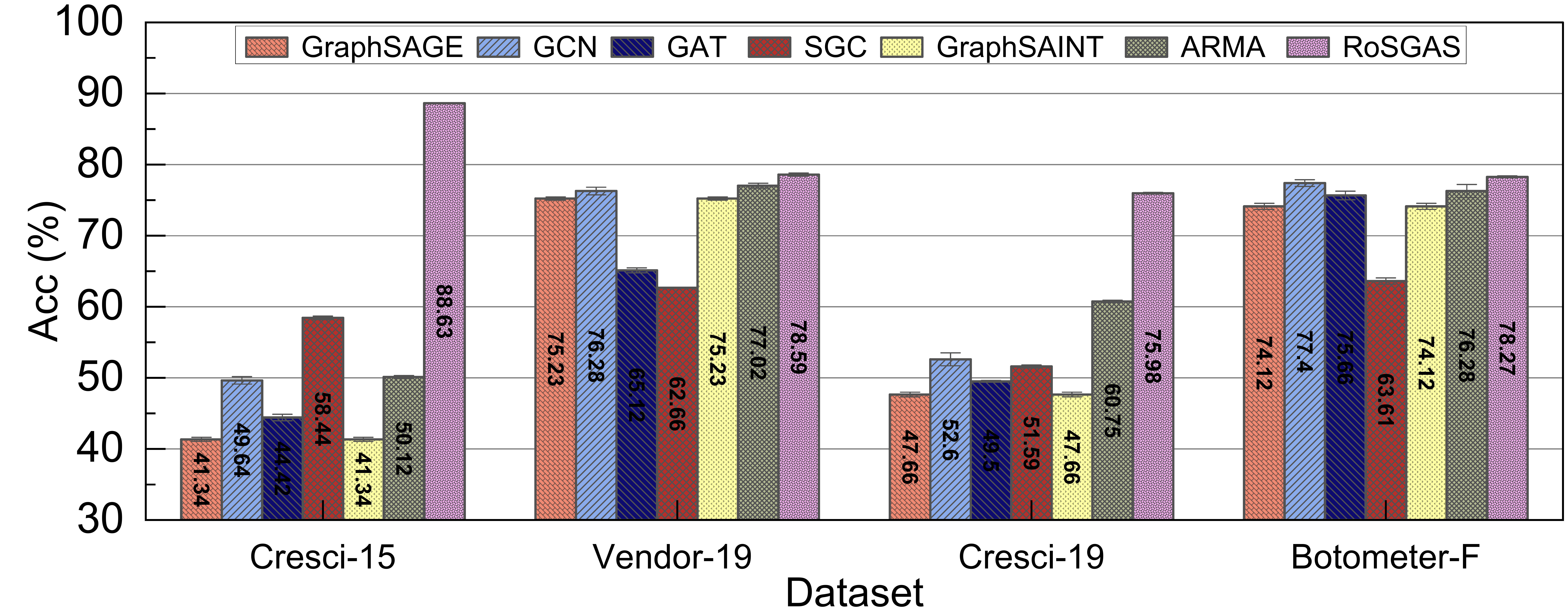}
\caption{Training on the Varol-17 and testing on different dataset}
\label{fig:general-varol-17}
\end{figure}

\begin{figure}[t]
\centering
\includegraphics[width=0.97\textwidth]{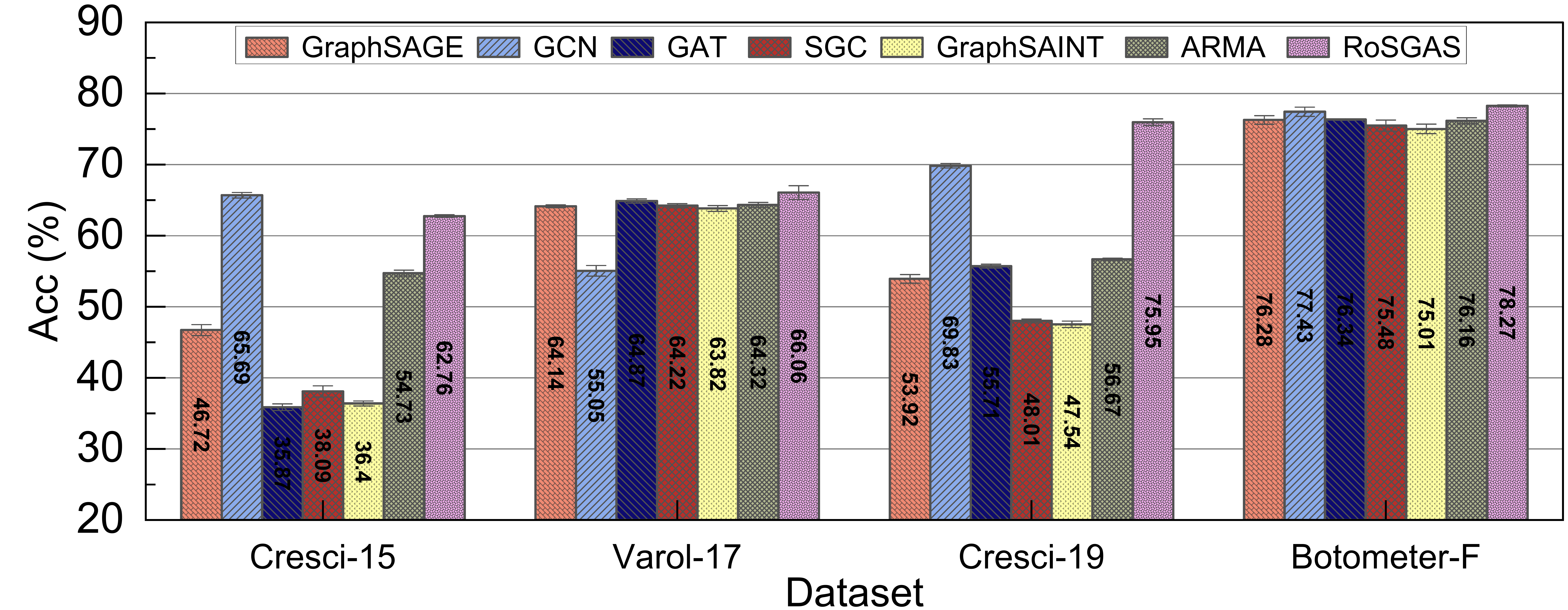}
\caption{Train on the Vendor-19 and testing on different dataset}
\label{fig:general-vendor-19}
\end{figure}

\begin{figure}[t]
\centering
\includegraphics[width=0.97\textwidth]{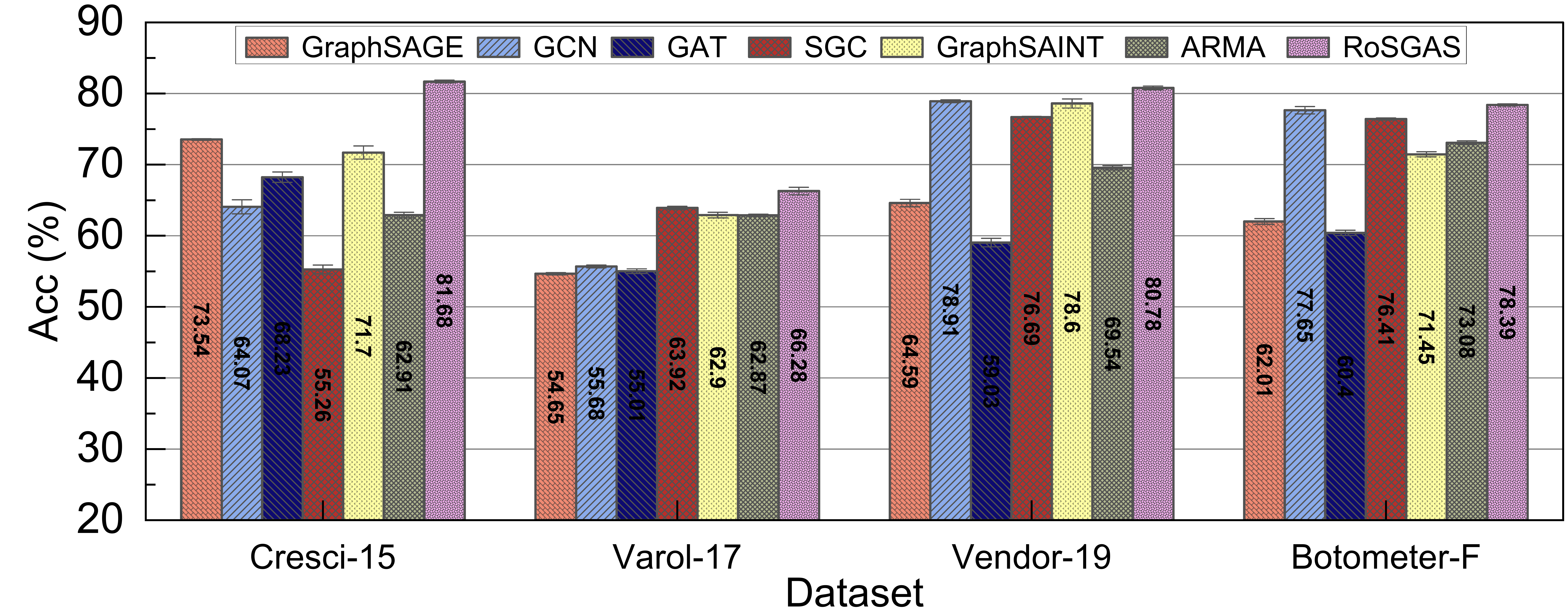}
\caption{Train on the Cresci-19 and testing on different dataset}
\label{fig:general-cresic-19}
\end{figure}

\begin{figure}[t]
\centering
\includegraphics[width=0.97\textwidth]{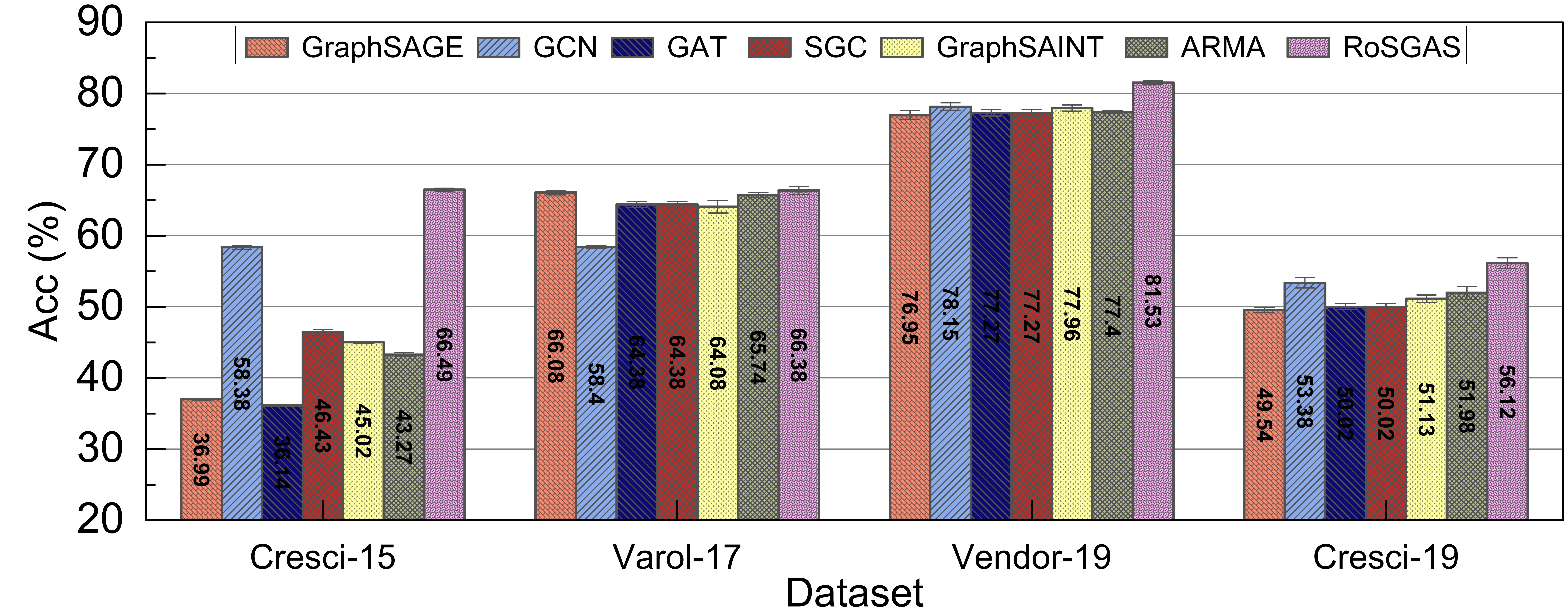}
\caption{Train on the Botometer-Feedback and testing on different dataset}
\label{fig:general-botometer-feedback}
\end{figure}

A series of figures including Fig.~\ref{fig:general-cresic-15}, Fig.~\ref{fig:general-varol-17}, Fig.~\ref{fig:general-vendor-19}, Fig.~\ref{fig:general-cresic-19} and Fig.~\ref{fig:general-botometer-feedback} show the prediction results when the training is based upon Cresci-15, Varol-17, Vendor-19, Cresci-19, and Botometer-Feedback, respectively. It is obviously observable that \RoSGAS outperforms all other baselines in the vast majority of scenarios. The only exceptions are when the model is trained based on Vendor-19 and validated on Cresci-15 dataset, or when the model is trained based on Cresci-15 whilst validating upon Vendor-19.  In this two cases, the accuracy of \RoSGAS is slightly lower than GCN. We reason this phenomenon is possibly because GCN is less sensitive to the disparity between two datasets. Unsurprisingly, all models has a reduced accuracy when conducting the out-of-sample prediction as opposed to its in-sample accuracy, simply because of the potential overfitting in the in-sample evaluations. The experiments carried out in this subsection generically showcase the robustness and generalization of our approach when handling new data where different noise manifests. 

\begin{figure}[hb]
\centering

\subfigure[GraphSAGE  (Score:$4.956\times10^{-1}$)]{
\centering
\includegraphics[width=0.3\textwidth]{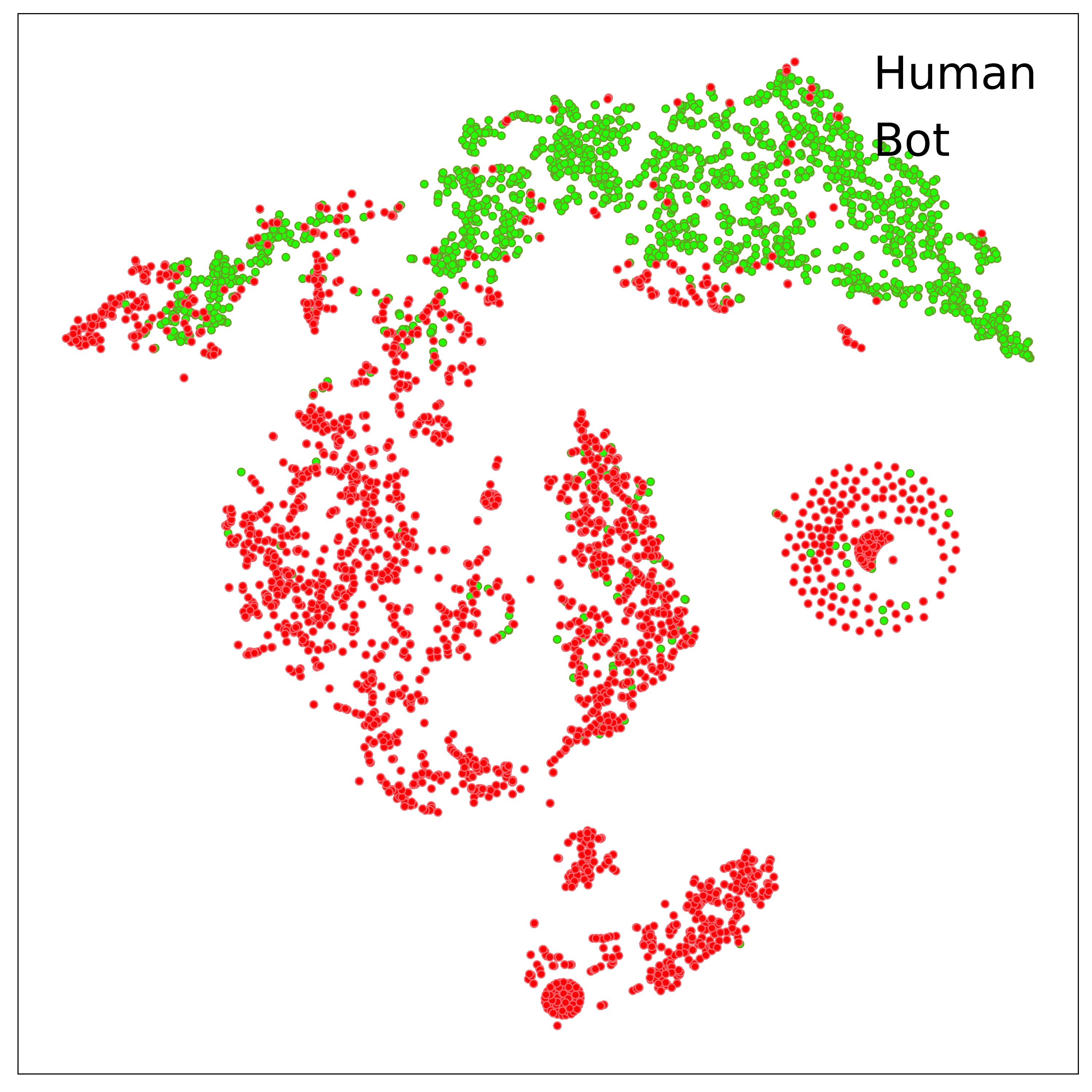}
}%
\quad
\subfigure[GraphSAGE (Score:$2.250\times10^{-1}$)]{
\centering
\includegraphics[width=0.3\textwidth]{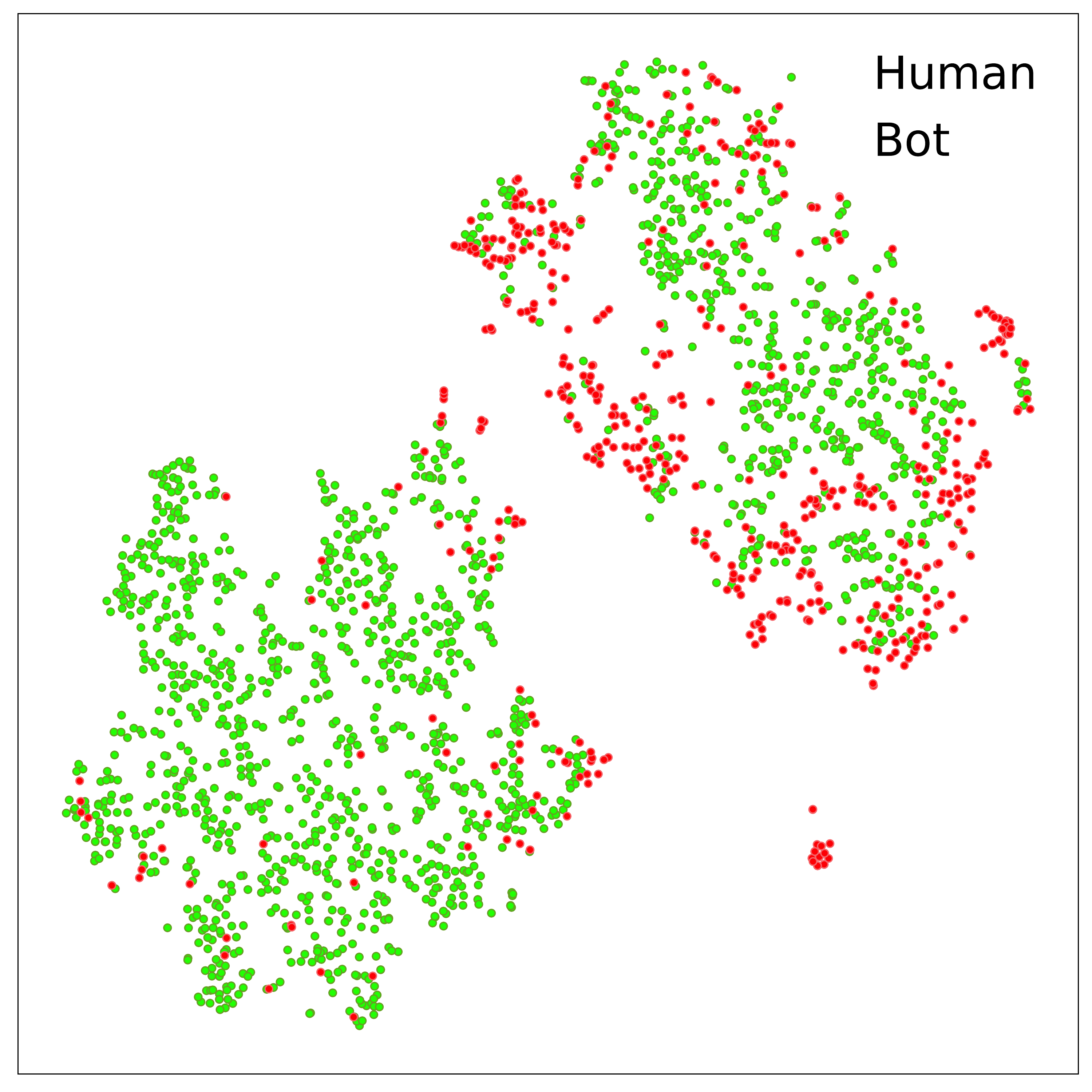}
}%
\quad
\subfigure[GraphSAGE  (Score:$1.825\times10^{-1}$)]{
\centering
\includegraphics[width=0.3\textwidth]{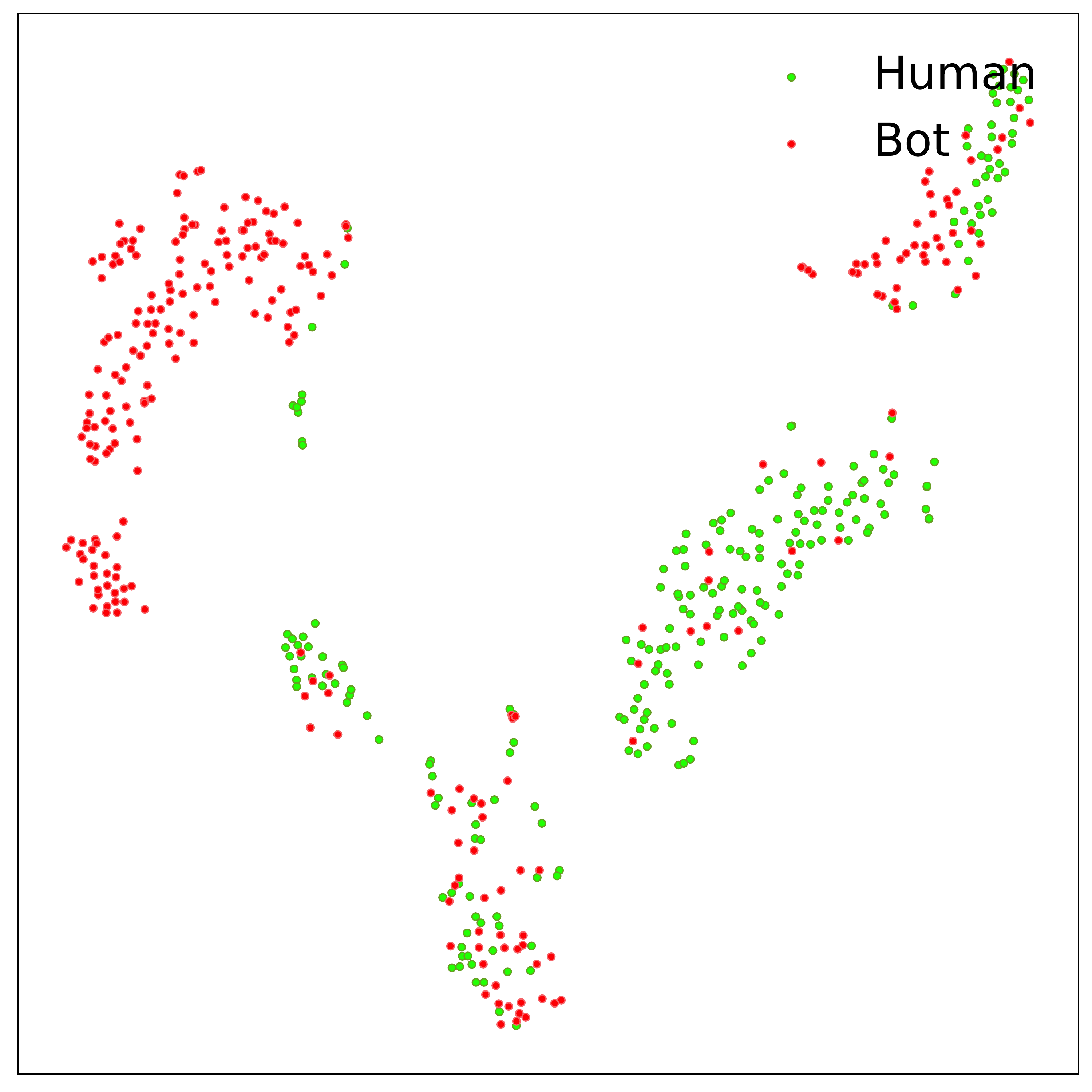}
}%

\subfigure[GCN  (Score:$1.101\times10^{-1}$)]{
\centering
\includegraphics[width=0.3\textwidth]{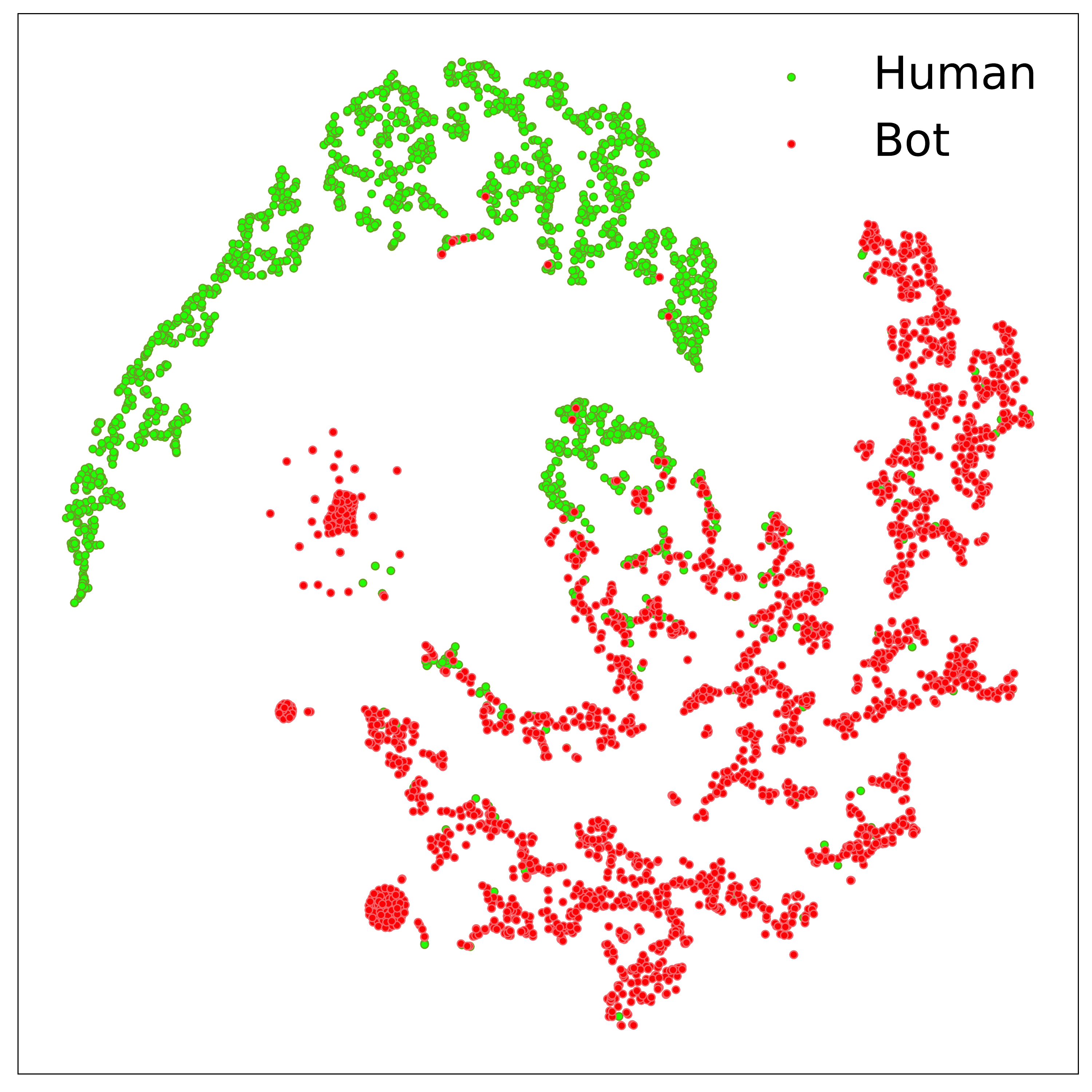}
}%
\quad
\subfigure[GCN  (Score:$4.158\times10^{-1}$)]{
\centering
\includegraphics[width=0.3\textwidth]{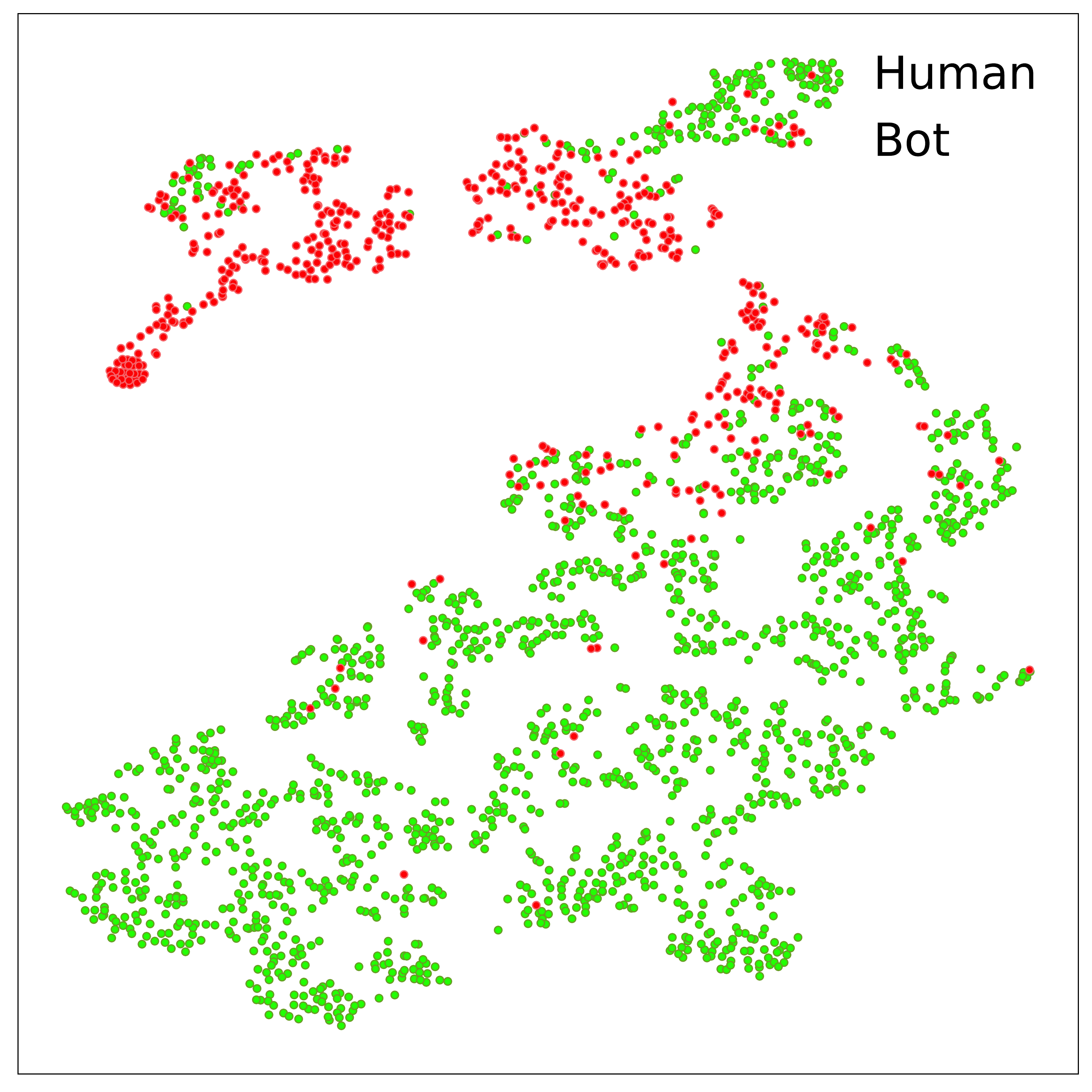}
}%
\quad
\subfigure[GCN (Score:$2.849\times10^{-1}$)]{
\centering
\includegraphics[width=0.3\textwidth]{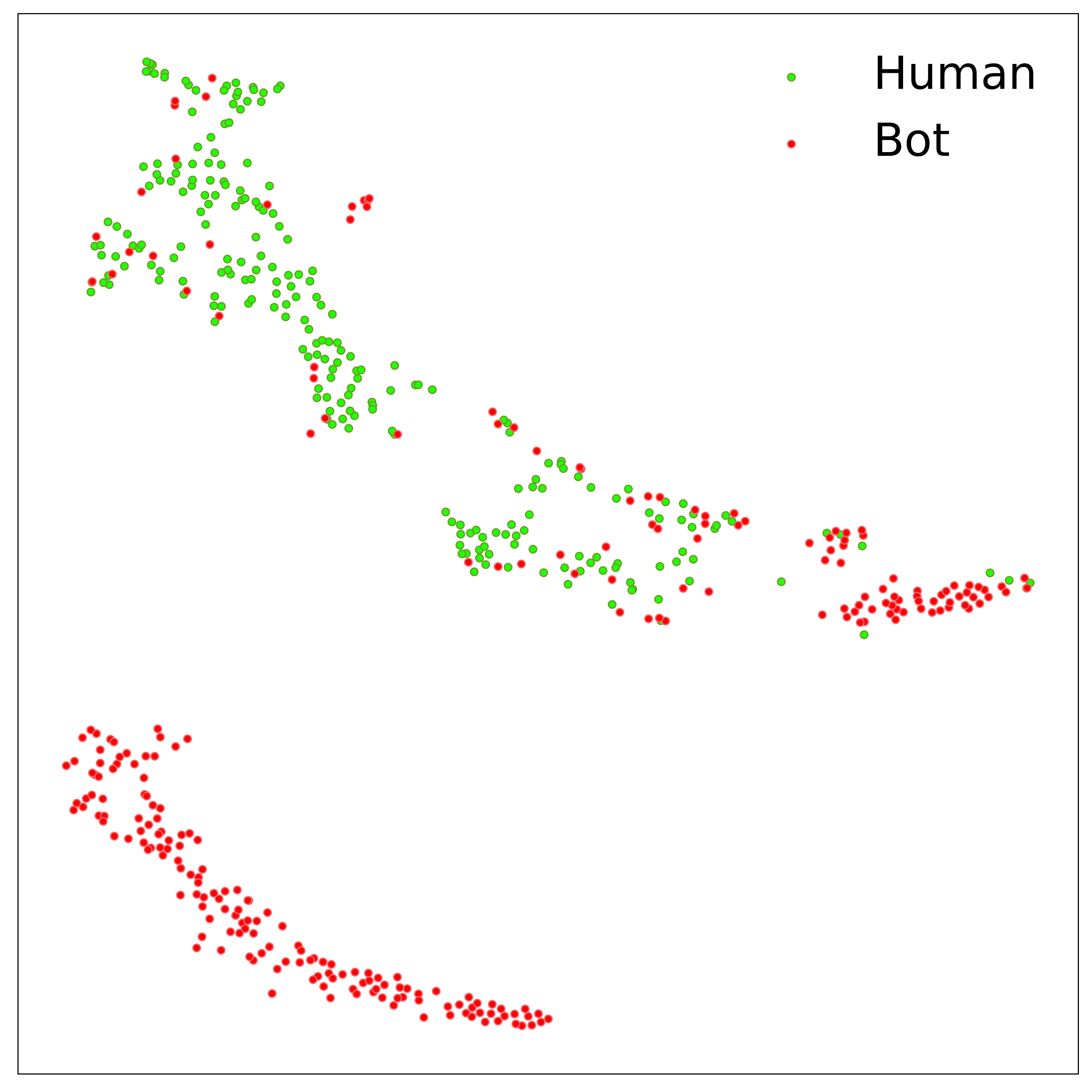}
}%
\quad
\subfigure[GraphSAINT (Score:$4.600\times10^{-1}$)]{
\centering
\includegraphics[width=0.3\textwidth]{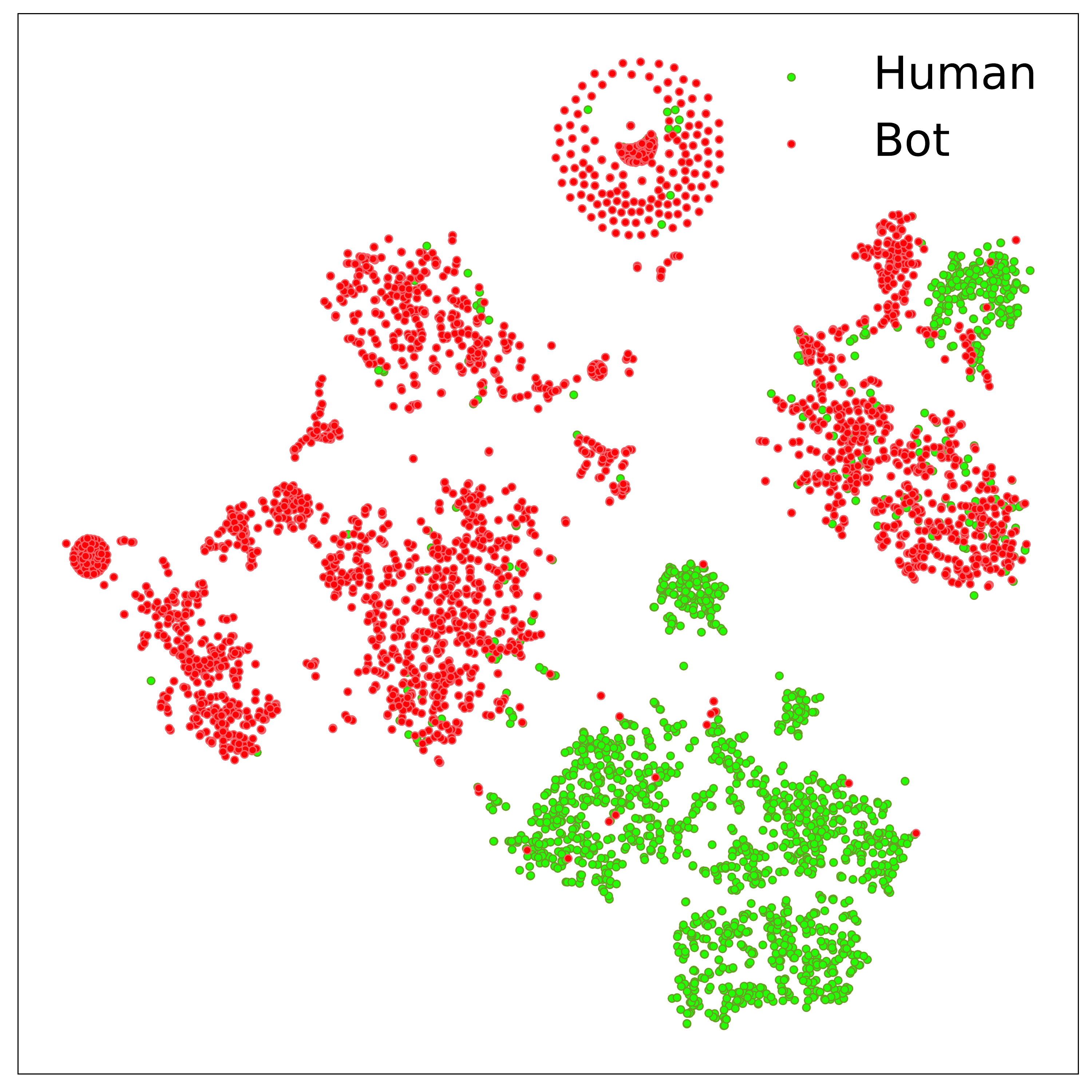}
}%
\quad
\subfigure[GraphSAINT  (Score:$1.799\times10^{-1}$)]{
\centering
\includegraphics[width=0.3\textwidth]{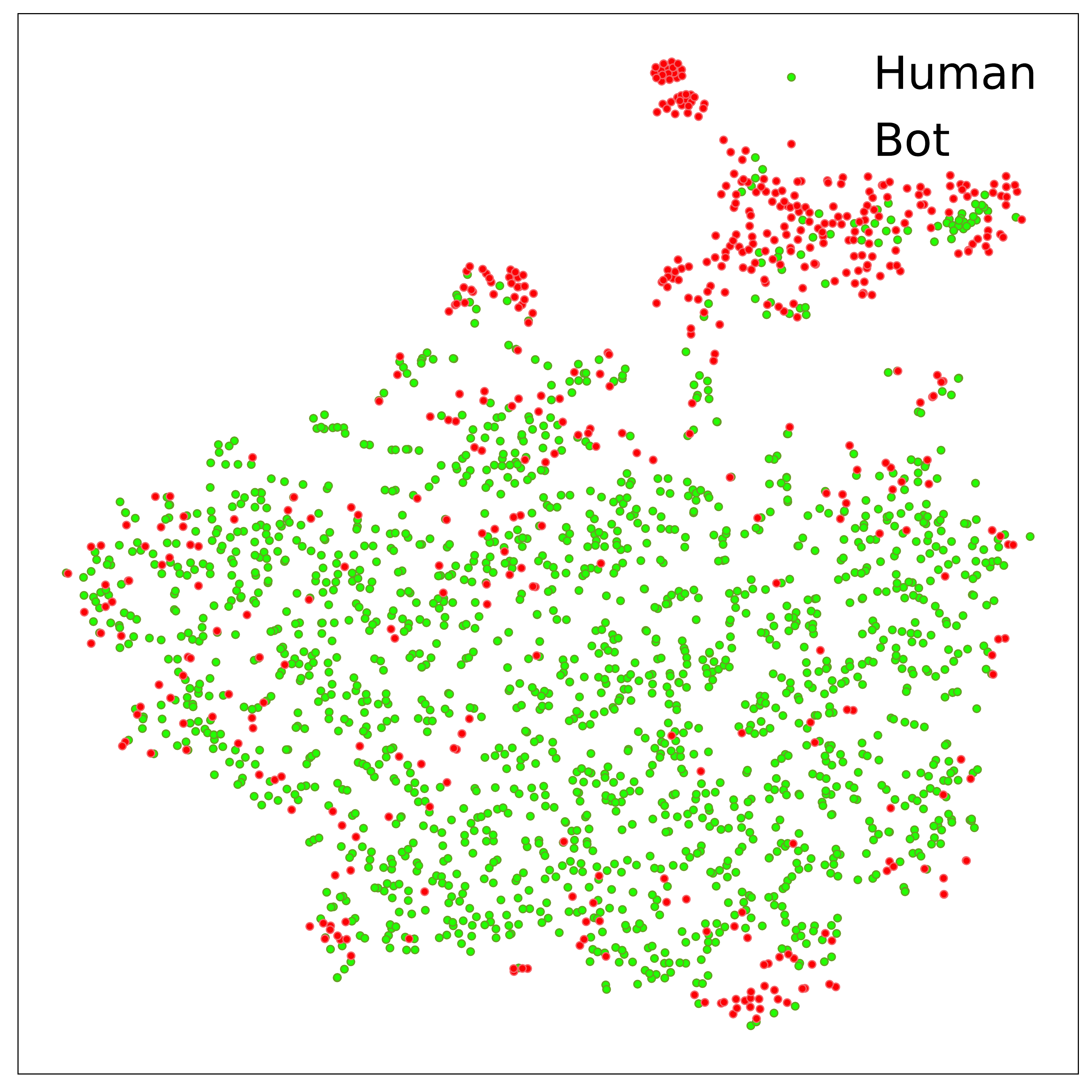}
}%
\quad
\subfigure[GraphSAINT (Score:$1.505\times10^{-1}$)]{
\centering
\includegraphics[width=0.3\textwidth]{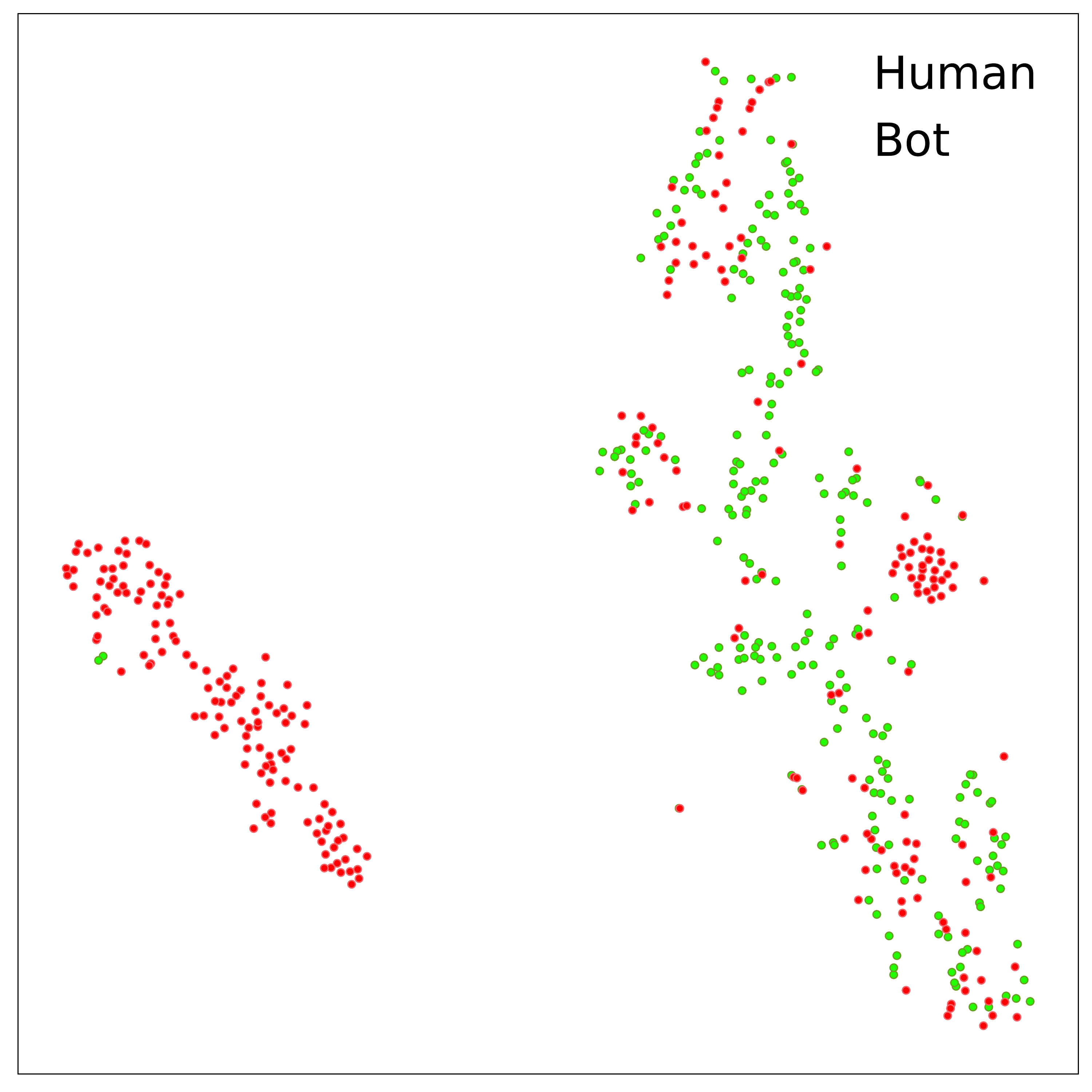}
}%
\end{figure}

\begin{figure}[h]

\subfigure[ARMA (Score:$9.448\times10^{-2}$)]{
\centering
\includegraphics[width=0.3\textwidth]{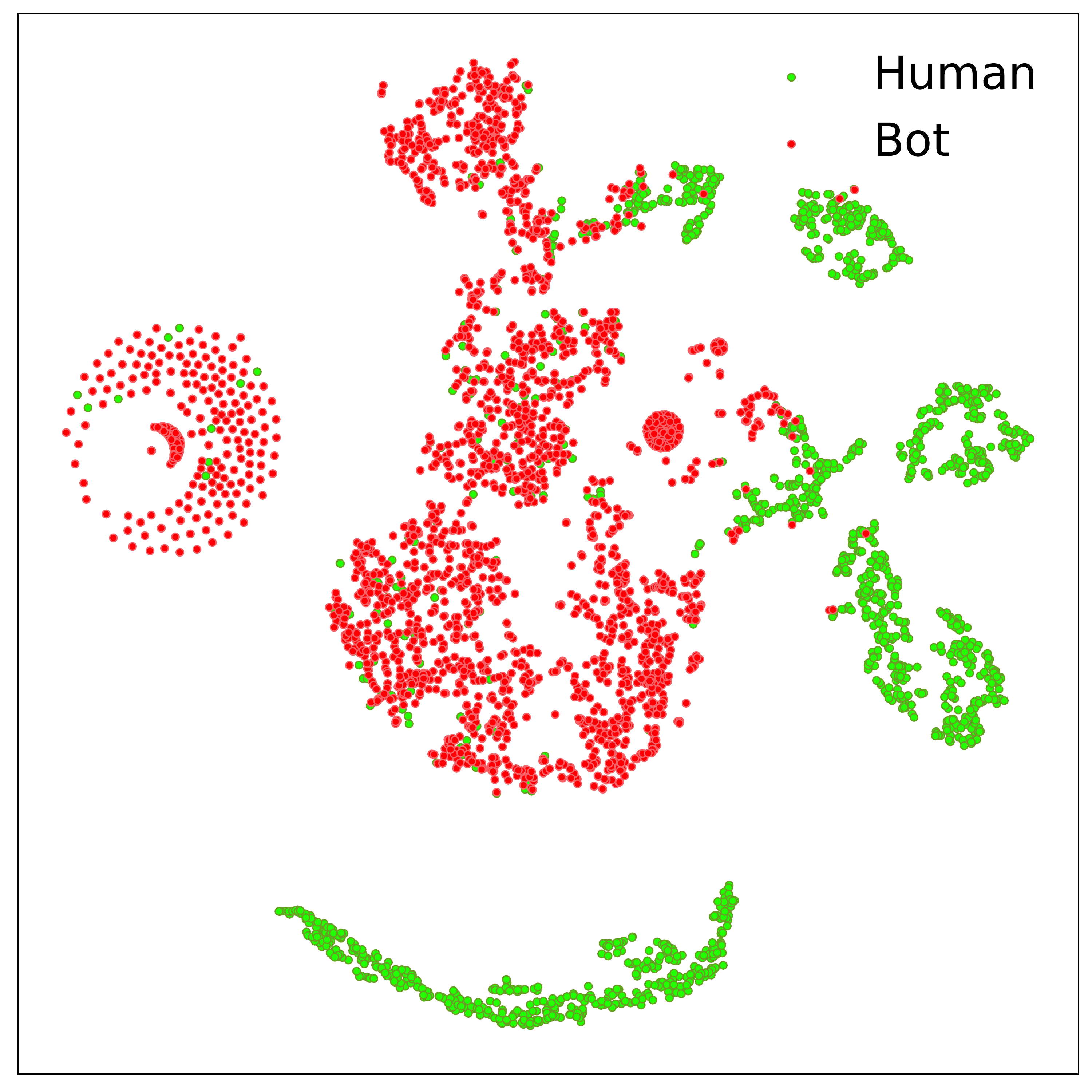}
}%
\quad
\subfigure[ARMA (Score:$3.223\times10^{-1}$)]{
\centering
\includegraphics[width=0.3\textwidth]{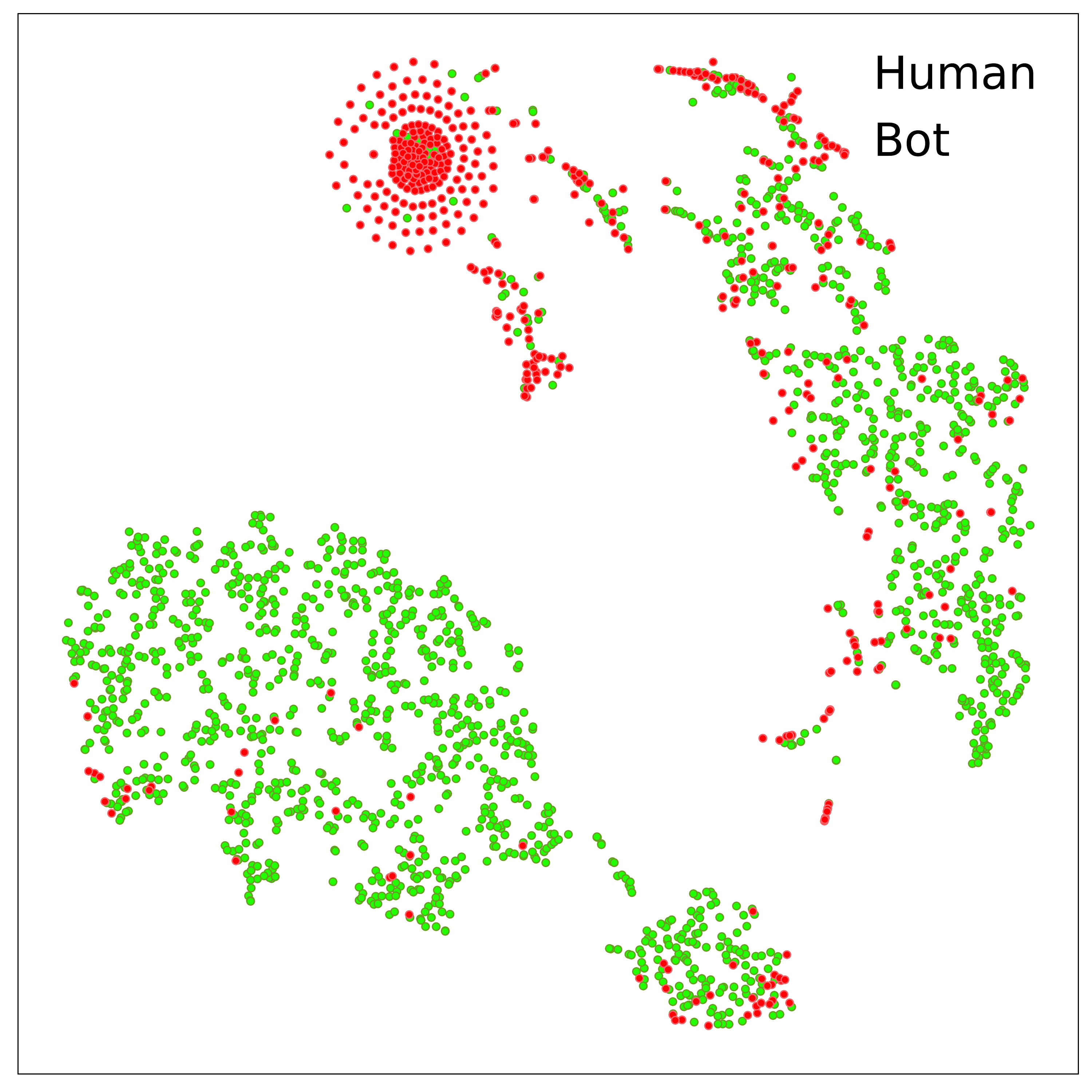}
}%
\quad
\subfigure[ARMA (Score:$1.847\times10^{-1}$)]{
\centering
\includegraphics[width=0.3\textwidth]{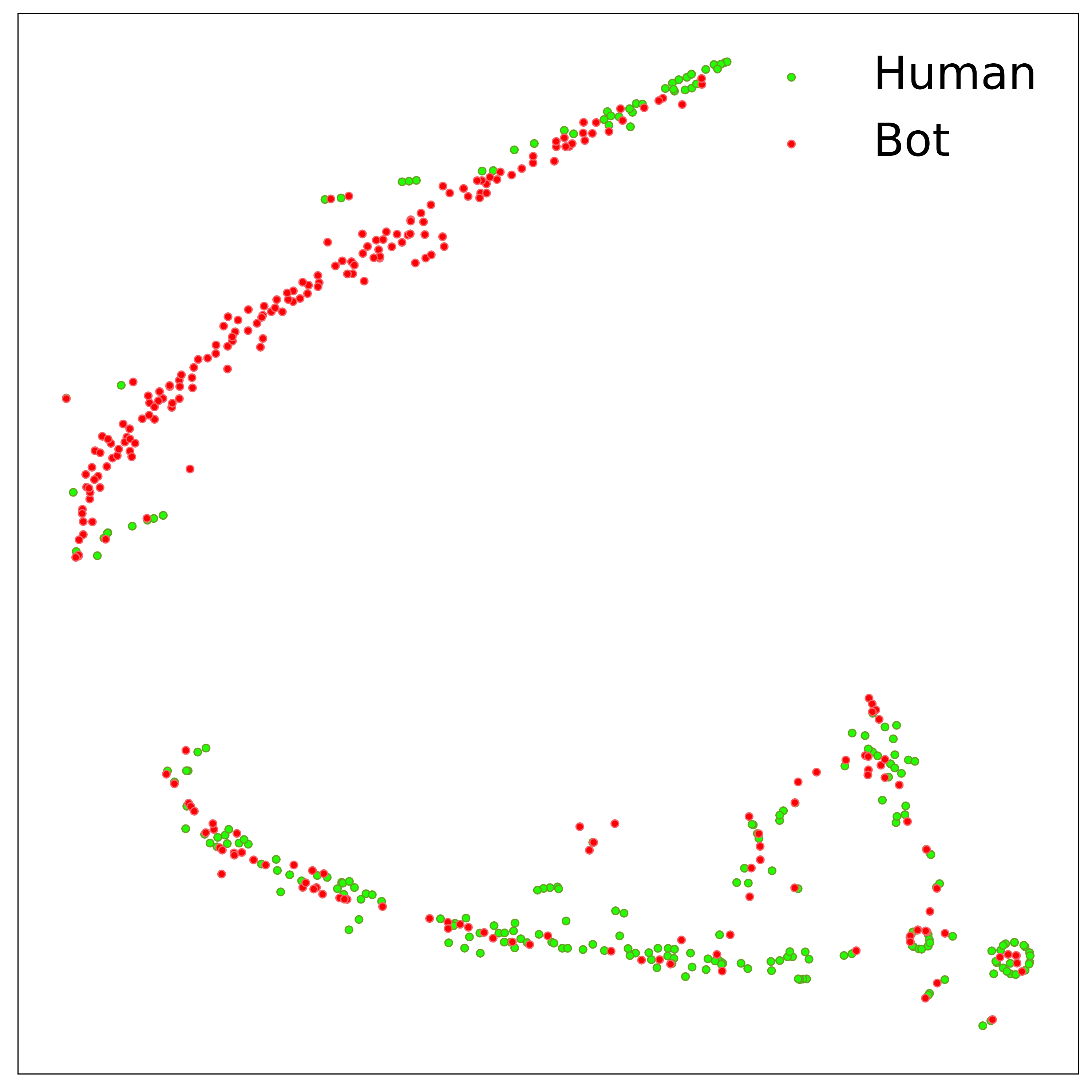}
}%

\subfigure[\RoSGAS (Score:$8.174\times10^{-1}$)]{
\centering
\includegraphics[width=0.3\textwidth]{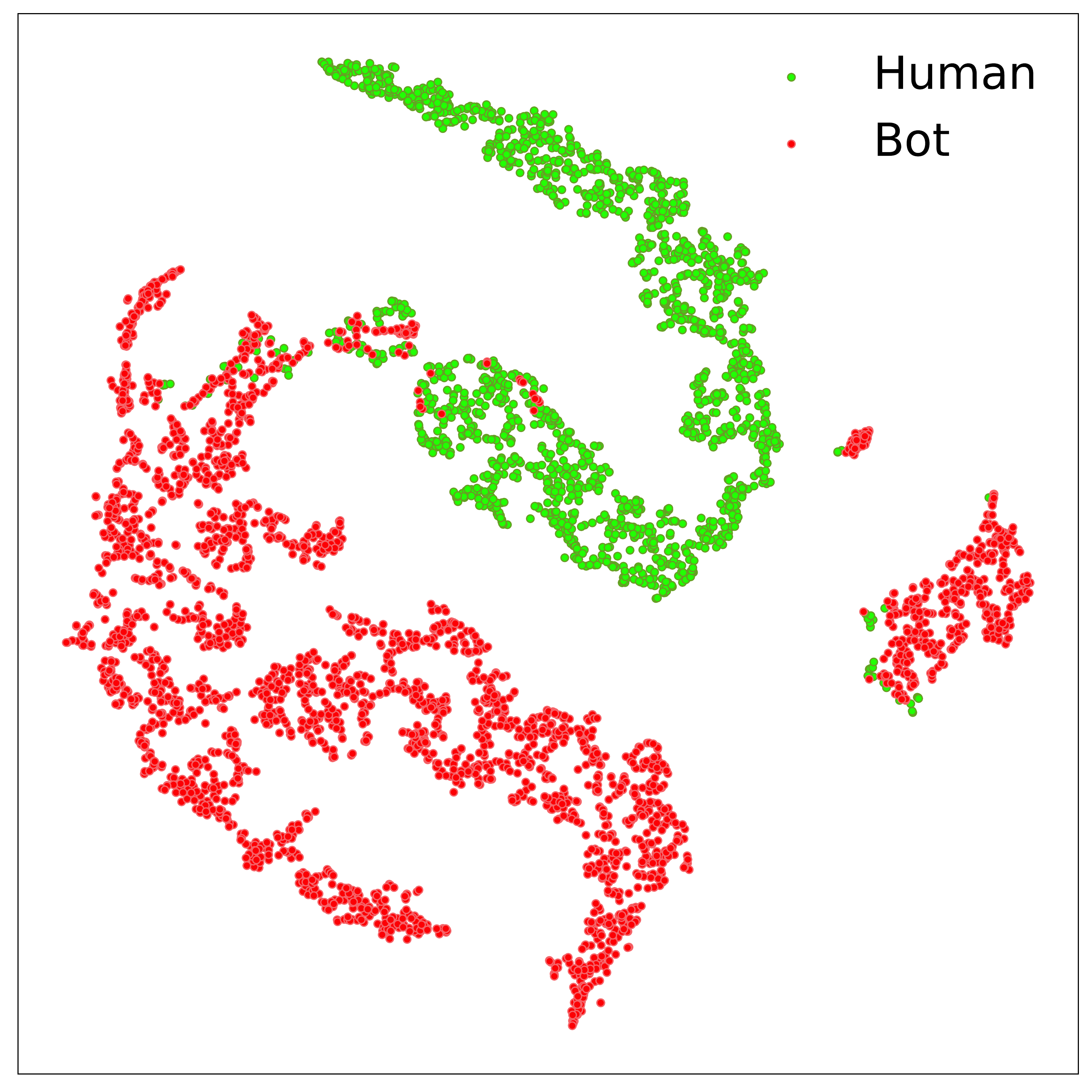}
}%
\quad
\subfigure[\RoSGAS (Score:$4.619\times10^{-1}$)]{
\centering
\includegraphics[width=0.3\textwidth]{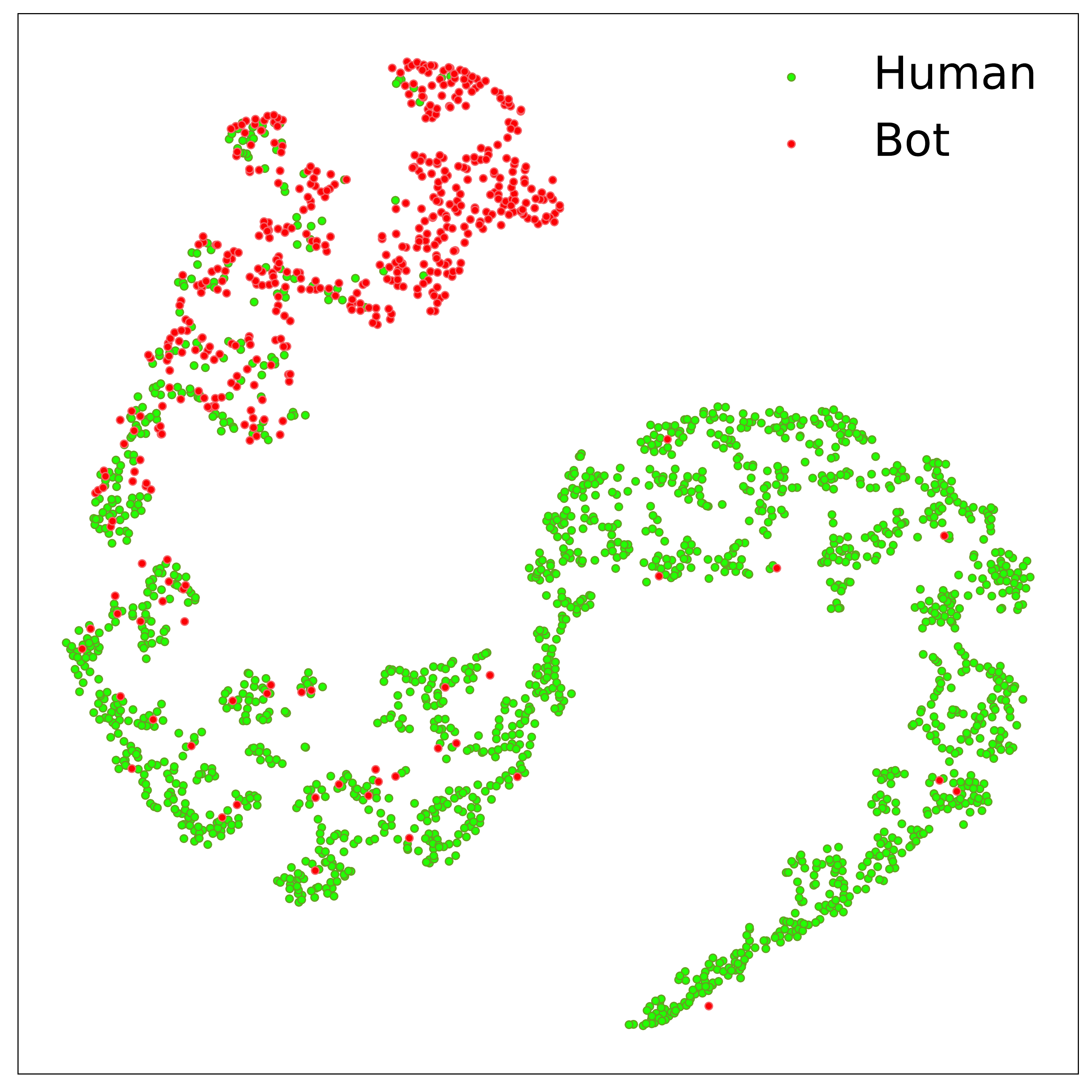}
}%
\quad
\subfigure[\RoSGAS (Score:$3.654\times10^{-1}$)]{
\centering
\includegraphics[width=0.3\textwidth]{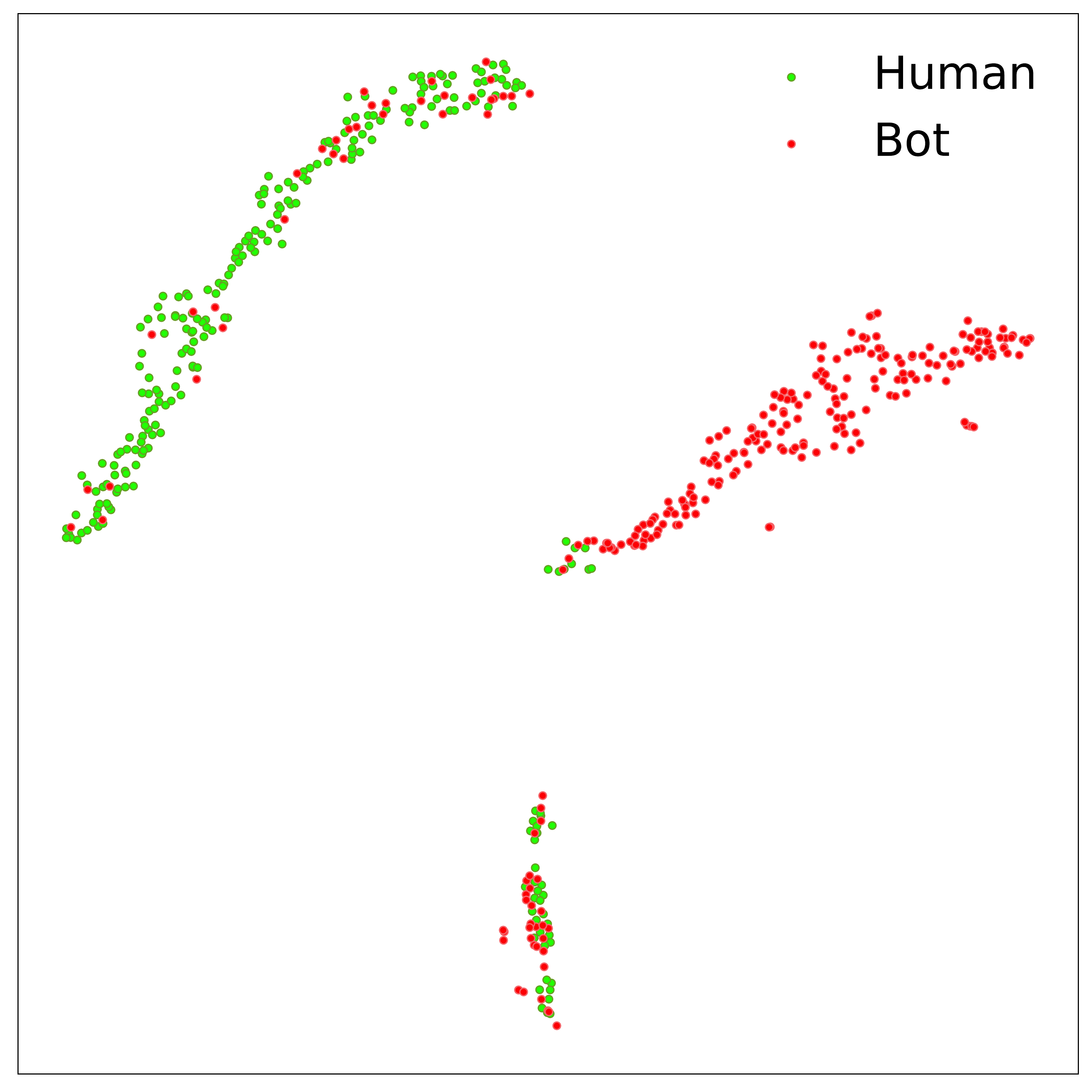}
}%

\centering
\caption{2D t-SNE plot of representation vectors of users produced by GraphSAGE, GCN, GraphSAINT ARMA and \RoSGAS on the dataset of Cresci-15 (left column), Vendor-19 (middle column) and Cresci-19 (right column), respectively. The corresponding homogeneity score is given in the bracket}
\label{fig:representation_research}
\end{figure}

\subsubsection{Stability}

We can also observe the minimum performance fluctuation of \RoSGAS, compared with other baselines, when different datasets are used as test sets. On the contrary, many other baselines such as GraphSAGE, GCN, and GraphSAINT have a noticeable fluctuation. For example, as demonstrated in Fig.~\ref{fig:general-cresic-15}, when training on Cresci-15, the accuracy of GraphSAGE is only 23.21\% and 26.63\% when the trained model is tested based on Vendor-19 and Botometer-F. However, the accuracy can climb up to 47.8\% if tesing on Cresci-19. Likewise, as shown in Fig.~\ref{fig:general-varol-17}, while the accuracy of the GraphSAINT model trained upon Varol-17 and tested upon Cresci-15 is merely 41.34\%, a competitive accuracy can be obtained when the model is tested on Vendor-19 (75.23\%) or Botometer-F (74.12\%). 

The stable generalization stems from the adaptability and robustness of \RoSGAS. In fact, our method only exploits some common features for the task of detecting social bot, without tightly coupling with, or depending upon exclusive features or numerical characteristics. This will help maintain an outstanding quality of detection even when the testset varies.

\subsection{Case Study: Effectiveness of Representation Learning}
\label{sec:exp_res:casestudy}

To further understand the quality of vector representation, we compare the representations of \RoSGAS with the baselines that can achieved good results in the accuracy evaluation, i.e., GraphSAGE, GCN, GraphSAINT and ARMA. For each individual model, we cluster the representation results by using $k$-means with $k=2$ and then calculate the homogeneity score -- a higher-the-better indicator that measures how much the sample in a cluster are similar. The homogeneity is satisfied -- the value equals 1 -- if all of its clusters contain only data points which are members of a single class.
 
Fig.~\ref{fig:representation_research} visualizes the result of t-SNE dimensionality reduction on the representation vectors of users. We evaluate each model based on three given datasets, Cresci-15, Vendor-19, Cresci-19, respectively. For each baseline model, the results on the three datasets are placed in a row.  
Numerically, the homogeneity score of \RoSGAS is higher than other baselines when adopting all datasets. For example, the score of \RoSGAS is over 2 times higher than that of GraphSAGE on average, indicating the most distinguishable represent vectors can be obtained by our approach. The visualization completely aligns with the measurement.  Observably, the bots in \RoSGAS can be far more easily differentiated from the benign users.

\section{Discussion}
\label{sec:discussion}

\mypara{Significance of subgraph-based and RL-guided solution}. This work is to advance the development and application of GNNs in the field of social bot detection. The performance of features-based, statistics-based and deep learning based methods fades facing the evolving bot technologies. We applied the GNN to leverage the information of the neighborhood and relationship to counter the development of bots.
This work is non-trivial when tackling massive datasets with hundreds of thousands or even millions of nodes. The existing solutions to complex model architecture search is not well-suited for the problem of social bot detection at scale under study -- the data distribution becomes more non-IID due to the evolution of the bots, which complicates the model design with good generalization in practice. The extremely-large scale of the social network graph also leads to tremendously different user structures and necessitates adaptive detection of such structures with reasonable computational costs.
To this end, we proposed to search subgraphs to realize the reduction of graph scale, the simplification of model architectures, and the improvements of the detection performance. 

\mypara{Necessity of using RL.} The width ($k$) of the subgraph cannot be a common hyperparameter shared by all nodes and requires node-by-node personalization. In fact, social bots tend to randomly follow benign accounts as a disguise. The selection of $k$ primarily derives from our behavioural studies -- There are noticeably enough bots in the subgraph within 2th-order subgraph while benign nodes could be well recognized in the 1th order subgraph. Increasing the value would not incur additional performance gain but involve an overwhelming number of neighbors. Nevertheless, the range can be flexibly configured to adapt to any other scenarios and datasets. This necessitates the adoption of RL to facilitate the customized parameter search for each individual subgraph.

\mypara{Dealing with real-time streaming scenarios.} Critical challenges for on-the-fly bot detection encompass the need of message delivery, distributed event store and the capability of handling revolution and uncertainty of events and entities in the social network platform over time. This is particularly intricate due to the presence of new bots. The accuracy of offline learning models highly relies on the quality and quantity of the dataset that is fed into the models. However, this will be time-consuming and costly in real-time scenarios and the model update is required to maintain the high standard of model accuracy. 
In practice, to implement the streaming pipeline, a distributed crawler needs to be developed to continuously fetch social network information. The collected data is then forwarded to processing modules through distributed event streaming such as Kafka. More online and incremental designs are desired to underpin the on-the-fly version of the current bot detection framework.

\vspace{+0.4em}
\section{related work}
\label{sec:related}

In this section, we summarize the related literature and state-of-the-art approaches. The existing literature can be roughly classified into three categories: GNN based approaches, subgraph and RL based approaches, and self-supervised enhanced approaches.

\subsection{GNN based Social Bot Detection}
Early social bot detection mainly focuses on manually analyze the collected data to find discriminative features that can be used to detect the social bot.
However, the detection features are easy to imitate and escaped by social bots which are constantly evolving, and eventually become invalid.
The recent boom of Graph Neural Networks (GNNs) has promoted the latest progress in social bot detection.
The first attempt to use GNN to detect social bots \cite{ali2019detect} is mainly by combining the graph convolutional neural network with multilayer perception and belief propagation.
The heterogeneous graph is constructed and original node features are extracted by the pre-trained language model to get the final node embedding after aggregating by the R-GCN model \cite{feng2021botrgcn}, the BotRGCN successfully surpass the performance of traditional detection methods on the newly released dataset called TwiBot-20 \cite{feng2021twibot}. 
The heterogeneous graph is also applied in \cite{feng2021heterogeneity}, it also proposes a relational graph transformer inspired from natural language processing to model the influence between users and learn node representation for better social bot detection, and its performance exceeds the BotRGCN. 
However, these node embedding and node classification based methods perform convolution at the level of the entire graph, after stacking multiple layers of GNN on the tremendous scale of social graph will cause the over-smoothing problem \cite{li2018deeper,xu2018representation}.
Subgraph provides a new perspective to solve this issue, \cite{liu2018heterogeneous} constructs a heterogeneous graph and reconstructs subgraphs based on manually-defined heuristic rules for detecting malicious accounts in online platforms.
However, this kind of manual-based method not only consumes energy to set the extraction rules but also cannot be easily generalized to the field of social bot detection.

\subsection{Subgraph and RL based Approaches}
The scale of the graph constructed from the social network is tremendous. Performing convolution operation on the entire graph will not only consume computing resources but also lead to performance degradation due to problems such as over-smoothing.  The extraction of subgraphs either requires domain-specific expert knowledge to set the heuristic rules \cite{liu2018heterogeneous,yuan2020phishing} or needs to design motif for matching subgraphs \cite{yang2018node,peng2020motif}, which drastically limits the ﬂexibility and capability of generalization.
To address this issue, we leverage Reinforcement Learning (RL) to adaptively extract subgraphs. There have been a few attempts to marry RL and GNNs.  DeepPath \cite{xiong2017deeppath} establishes a knowledge graph embedding and reasoning framework that utilizes the RL agent to ascertain the reasoning paths in the knowledge base.
RL-HGNN \cite{zhong2020reinforcement} devises different meta-paths for any node in a HIN for adaptive select meta-path to learn its effective representations.
CARE-GNN \cite{dou2020enhancing}, RioGNN \cite{peng2021reinforced} RTGNN \cite{zhao2022multiview} and FinEvent \cite{peng2022event} all marry the RL and GNNs for dynamically optimizing the similarity threshold to achieve the purpose of selecting more valuable neighbour nodes for the aggregated nodes, to obtain more effective representation vectors for fraud detection or event detection.
Policy-GNN \cite{lai2020policy} utilizes RL to select the number of GNN architecture layers for aggregating the node embedding vectors to classify nodes.
GraphNAS \cite{gao2020graph} is among the first attempts to combine the recurrent neural network with GNNs. It continuously generates descriptions of GNN architecture to find the optimal network architecture based on RL by maximizing the expected accuracy.
Similarly to the architecture search in GraphNAS, Auto-GNN \cite{zhou2019auto} additionally proposes a parameter sharing mechanism for sharing the parameters in homogeneous architecture for reducing the computation cost.
However, these methods are not combined with the subgraph method and their optimization is tightly coupled with  speciﬁc datasets. 
They are not suited for detecting graphs that follow a power-law distribution with huge disparities among different users \cite{barabasi1999emergence,muchnik2013origins}. By contrast, our approach relies on subgraph embedding to achieve high detection effectiveness without compromising time efficiency, and has strong generalization across multiple datasets. 

\subsection{Self-Supervised Learning Approaches on Graphs}

In recent years, self-supervised learning has hugely advanced as a promising approach to overcome the limited data annotation and limited and to enable a target objective  achieved without supervision. The technology has been investigated in a wide range of domains, such as natural language processing \cite{devlin2018bert,zhang2019hibert}, computer vision \cite{larsson2016learning,oord2016conditional} and graph analysis \cite{grover2016node2vec,kipf2016variational}. At the core of self-supervised learning is to define an annotation-free pretext task to train an encoder for representation learning. Particularly for graph analysis, there are a few works of literature about designing self-supervised tasks based on either edge attributes \cite{dai2018adversarial,tang2015line} or node attributes \cite{ding2018semi}. However, the inherent dependencies among different nodes in the topology hinder the appropriate design of the pretext tasks.
DGI \cite{velickovic2019deep} trains a node encoder to maximize the mutual information between the node representations and the global graph representation. 
GMI \cite{peng2020graph} defines a pretext task that maximizes the mutual information between the hidden representation of each node and the original features of its 1-hop neighbors.
InfoGraph \cite{sun2019infograph} maximizes the mutual information between the graph embeddings and the substructure embeddings at different scales to more effective learn graph embeddings. 
However, the aforementioned self-supervised learning approaches need to take the holistic graph as the input, which is time- and resource- consuming and thus restricts the scalability on large-scale graphs. Our approach aims to obtain a subgraph-level representation to ensure non-homologous subgraphs are discriminative while homologous subgraphs have similar representation vectors. 

\section{conclusion}
\label{sec:conclusion}

This paper studies a RL-enabled framework for GNN architecture search. The proposed  \RoSGAS framework can adaptively ascertain the most suitable multi-hop neighborhood and the number of layers in the GNN architecture when performing the subgraph embedding for the social bot detection task. We exploit HIN to represent the user connectivity and use multi-agent deep RL mechanism for steering the key parameter search. The subgraph embedding for a targeted user can be more effectively learnt and used for the downstream classification with competitive accuracy whilst maintaining high computation efficiency. Experiments show that \RoSGAS outperforms the state-of-the-art GNN models in terms of accuracy, training efficiency and stability. \RoSGAS can more quickly achieve, and carry on with, high accuracy during the RL training, and has strong generalization and explainability. We believe the data-centric solution guided by behavioural characterization and reinforcement learning -- instead of heavily complicating the network architecture itself -- would be a promising and innovative direction, which is both scientifically and engineering-wise challenging in the field of bot detection. In the future, we plan to examine the impact of feature distribution and graph structure on model training, and extend \RoSGAS to underpin the streaming scenarios.

\section*{Acknowledgment}
We thank anonymous reviewers for the provided helpful comments on earlier drafts of the manuscript. 
Zhiqin Yang and Yue Wang are supported by the National Key R\&D Program of China through grant 2021YFB1714800, and S\&T Program of Hebei through grant 20310101D.
Yangyang Li is supported by NSFC through grant U20B2053.
Yingguang Yang, Kai Cui and Haiyong Xie are supported by National Key R\&D Program of China through grant SQ2021YFC3300088.
Renyu Yang and Jie Xu are supported by UK EPSRC Grant (EP/T01461X/1), UK Turing Pilot Project funded by the Alan Turing Institute. Renyu Yang is also supported by the UK Alan Turing PDEA Scheme.

\bibliographystyle{ACM-Reference-Format}
\bibliography{main}

\end{document}